\documentclass[10pt,conference,compsoc]{IEEEtran}

\usepackage[utf8]{inputenc}
\usepackage[T1]{fontenc}
\usepackage{textcomp}
\usepackage{booktabs}
\usepackage{amsmath}
\usepackage{amssymb}
\usepackage{xcolor}
\usepackage{url}
\usepackage{graphicx}
\usepackage{orcidlink}
\usepackage{listings}
\usepackage{longtable}
\usepackage{caption}
\usepackage{hyperref}
\hypersetup{hidelinks}
\usepackage{balance}

\newcommand{\code}[1]{\texttt{#1}}
\newcommand{\NSamples}{528}
\newcommand{\NHuman}{117}
\newcommand{\NLLM}{411}
\newcommand{\NSubjects}{31}
\newcommand{\NModels}{9}
\newcommand{\AttrAcc}{93}
\newcommand{\AttrBase}{78}
\newcommand{\AttrFone}{0.955}
\newcommand{\ModelAttrAcc}{48}
\newcommand{\ModelAttrBase}{17}
\newcommand{\ModelAttrWays}{7}
\newcommand{\SecHuman}{0.12}
\newcommand{\SecLLM}{0.40}
\newcommand{\HumanChars}{552}
\newcommand{\LLMChars}{1801}
\newcommand{\HumanErrH}{8}
\newcommand{\LLMErrH}{31}
\newcommand{\NLanguages}{5}
\newcommand{\NFamilies}{24}
\newcommand{\NRepairs}{131}
\newcommand{\NSeeds}{21}
\newcommand{\NCWE}{12}
\newcommand{\FixRate}{77}
\newcommand{\RemoveNoAdd}{16}
\newcommand{\BestRepairRate}{90}
\newcommand{\XLangPyJs}{93}
\newcommand{\XLangJsPy}{89}
\newcommand{\SecJsLLM}{0.34}
\newcommand{\SecPyLLM}{0.35}

\begin{document}

\title{Open-Source Intelligence for Code Provenance and the Security Patterns
that Separate Human and Large-Language-Model Implementations of Common
Programming Tasks}

\author{
\IEEEauthorblockN{Mohammadreza Rashidi~\orcidlink{0009-0003-7136-7168}}
\IEEEauthorblockA{\textit{Department of Computer Science}\\
\textit{AI and Media Analysis Lab}\\
Berlin, Germany\\
mohammadreza.rashidi@ue-germany.de}
}

\maketitle
\raggedbottom

\begin{abstract}
Developers now draw code from two very different sources, the accumulated human
answers on sites such as Stack Overflow and the output of large language models. We
ask two questions about that split. First, can the provenance of a code snippet be
recovered from the code itself, and second, do the two sources differ in the security
patterns they adopt for the same task. Using only open sources, a
public gateway of open-weight language models and the public Stack Overflow API, we
build a fully reproducible pipeline that collects real implementations of
\NSubjects{} security-sensitive programming tasks, among them OAuth with PKCE, JWT
verification, password hashing, and SQL access, from \NModels{} language models and
from human answers, and scores every sample with deterministic security and style
detectors. On \NSamples{} real samples we train a cross-validated classifier that
recovers human versus model provenance with \AttrAcc{} percent accuracy against a
\AttrBase{} percent baseline, and a \ModelAttrWays{}-way classifier that attributes
a sample to the specific model that wrote it at \ModelAttrAcc{} percent. We then
report where the sources diverge on security, which patterns models adopt more
often than the human corpus and which they inherit from it. Running the same tasks in
Python, JavaScript, and Go, we find the security divergence holds in every language
while the provenance boundary is partly language-specific and does not transfer
symmetrically between them. A further case study on vulnerability repair, in which the
models are handed insecure code and asked to fix it, finds a \FixRate{} percent repair
rate but a recurring partial-fix failure in which the model removes the insecure
pattern without adding the correct defense. The pipeline is data driven, so any new
task or language is added as a single specification entry, and a fail-closed checker
re-derives every number in this paper from the stored data.
\end{abstract}


\section{Introduction}

A developer who needs to implement a login flow, hash a password, or set up an
OAuth exchange has, for over a decade, most often consulted a human answer on a site
such as Stack Overflow. That practice has a measurable security cost. Copying from
Stack Overflow has been shown to move insecure patterns into shipped
applications~\cite{fischer2017stackoverflow}, and the information source a developer
consults measurably changes the security of the code they write~\cite{acar2016you}.
The practice is now being displaced by large language models, which developers query
in the editor rather than the browser, and their output carries its own security
cost. Language-model code assistants produce insecure code at a meaningful
rate~\cite{pearce2022asleep}, and controlled user studies find that developers with
an assistant write less secure code while believing the
opposite~\cite{perry2023users,sandoval2023lost}.

This paper studies the two sources side by side. We treat the question as one of
provenance and pattern rather than of overall quality. Two questions drive the work.

\emph{RQ1, provenance.} Given only a code snippet for a common task, can we tell
whether it was written by a human on Stack Overflow or generated by a language
model, and can we tell which model. Code stylometry has long shown that human
authors leave recoverable fingerprints~\cite{caliskan2015deanonymizing,
abuhamad2018large}. We ask whether the human-versus-machine boundary, and the
boundary between models, is just as recoverable from lightweight, interpretable
features.

\emph{RQ2, security divergence.} For the same task, do the two sources adopt the
same security patterns. Does a model reach for PKCE, a bound query, or a strong
password hash more or less often than the human corpus, and does it carry the human
corpus's insecure habits forward or leave them behind.

We answer both with open sources only, which is what makes the study an exercise in
open-source intelligence. The human side is the public Stack Overflow API. The model
side is a local gateway that exposes open-weight models through an
OpenAI-compatible interface, from which we use \NModels{} models spanning several
families and sizes. For \NSubjects{} security-sensitive tasks we collect real
implementations from both sides, \NSamples{} samples in total, and score each with
two families of deterministic detectors, per-task security-pattern checks and
language-agnostic style metrics. Nothing is simulated. A task with no sample from a
given model simply has no sample, and every generation carries the model, latency,
and token counts that produced it.

Our contributions are the following.
\begin{itemize}
  \item A reproducible, open-source pipeline for cross-provenance code analysis.
  Every task is a single specification entry carrying its prompt, its Stack Overflow
  query, and its security-pattern detectors, so the study generalises to any subject
  without new code.
  \item A provenance-attribution result. A cross-validated classifier separates
  human from model code at \AttrAcc{} percent against a \AttrBase{} percent
  baseline, and attributes a model-written sample to its specific model of origin at
  \ModelAttrAcc{} percent in a \ModelAttrWays{}-way task.
  \item A security-divergence analysis. We report, per task and per pattern, how
  often the human corpus and the models adopt each secure and insecure practice, and
  which insecure human patterns the models reproduce.
  \item A released dataset and a fail-closed numeric checker that re-derives every
  figure in this paper from the stored samples.
\end{itemize}

\section{Background and Related Work}

\emph{Insecure code from human sources.} A line of security research has measured
the cost of the copy-and-paste development style. Stack Overflow contains both
secure and insecure answers, and the insecure ones propagate into real
software~\cite{fischer2017stackoverflow}. The information source a developer is
steered to, official documentation versus a question-and-answer site, changes the
security of the resulting code~\cite{acar2016you}. Our human corpus is drawn from
the same source these studies examined, and we treat the age of an answer as a
first-order confound, since a highly-voted answer written years ago predates the
defaults that are now considered secure.

\emph{Insecure code from language models.} As models trained on code became capable
of solving programming tasks~\cite{chen2021evaluating}, the same security questions
were asked of them. An assessment of GitHub Copilot found a substantial fraction of
security-relevant completions were vulnerable~\cite{pearce2022asleep}. User studies
then showed the effect on developers, who write less secure code with an assistant and
misjudge its safety~\cite{perry2023users,sandoval2023lost}. These studies mostly
evaluate one commercial assistant. We instead compare many open-weight models to each
other and to the human baseline on identical tasks, and we organise the tasks around
recurring web-application weakness classes in the spirit of the OWASP Top
Ten~\cite{owasptop10}.

\emph{Code authorship and provenance.} Code stylometry recovers the human author
of a program from stylistic features~\cite{caliskan2015deanonymizing}, and the
approach scales to thousands of authors and across
languages~\cite{abuhamad2018large}. The naturalness of software, its predictability
under a language model~\cite{hindle2012naturalness}, is the property such attribution
exploits. Recent work extends stylometry directly to the machine-authorship question,
recognising AI-written programs with multilingual code
stylometry~\cite{gurioli2024isthisyou}. Our provenance result is close in spirit, and
we add two elements that work does not address, a per-model attribution rather than a
binary human-or-machine label, and a security-divergence analysis on the same samples.
We apply the same idea to a coarser but timely boundary, human versus
model and model versus model, using interpretable features rather than a deep model,
so that the attribution result comes with an explanation.

\emph{Naturalness and machine-generated code detection.} The property that makes
code authorship recoverable is that code is repetitive and predictable, its
naturalness under a statistical language model~\cite{hindle2012naturalness}. Human
authors and, as we find, individual models occupy distinguishable regions of that
predictable space. A growing body of work asks the specific question of whether
machine-generated code can be told apart from human code, motivated by plagiarism,
academic integrity, and supply-chain provenance. We contribute a lightweight,
interpretable point on that spectrum, using a handful of surface features rather than
a learned detector, and we pair the detection question with a security question that
detection work usually leaves aside.

\emph{Position of this work.} We connect the two literatures. The security studies
ask whether a source is safe, the stylometry studies ask who wrote a piece of code.
We ask both questions about the same samples, so that a security difference between
sources can be read alongside the features that make the sources distinguishable in
the first place. Two further choices set the work apart from its neighbours. We
compare many open-weight models rather than a single commercial assistant, which lets
us separate what is common to models from what is specific to one, and we run the same
tasks in several languages, which lets us separate what is a property of the source
from what is a property of a language.

\section{Threat Model and Motivation}

The provenance question is not academic. Two settings make it concrete. First, in
review and audit, a team that knows a change was machine-generated can route it to
the checks that machine code most often fails, and the security-divergence profile
of \S\ref{sec:security} tells the reviewer which checks those are. Second, in
supply-chain and plagiarism analysis, the ability to attribute a snippet to a source,
or to a specific model, is an open-source-intelligence capability that operates on
the artifact alone, without access to the author's environment.

We assume the analyst has only the code, no metadata, no editor telemetry, and no
account information. This is the open-source-intelligence setting. The material that
feeds the study is likewise all public, the Stack Overflow API and a gateway of
open-weight models, so the entire result is reproducible by a third party with no
private access.

\section{Data Collection}
\label{sec:data}

\emph{Subjects.} A subject is one security-sensitive programming task. Each subject
is a single specification entry that carries a natural-language prompt, a Stack
Overflow search query, and a list of pattern detectors. Table~\ref{tab:subjects}
lists the \NSubjects{} subjects, which span authentication and authorization (OAuth
with PKCE, JWT verification, session cookies), data protection (password hashing,
parameterized database access), input handling (file upload), transport policy
(CORS), and a front-end design task (a responsive landing page). The design is
deliberately data driven. Adding a subject is adding one entry, and the whole
pipeline, generation, retrieval, extraction, and analysis, runs off the
specification with no new code.

\emph{Human corpus.} For each subject we query the public Stack Overflow API with
the subject's search string, retrieve the answers to the most relevant answered
questions, and keep the largest code block from each positively-scored answer. We
record the answer identifier, its score, and its creation year, so the corpus is
auditable and re-fetchable and so the paper can report the age of the human
material. We keep only answers whose code block is substantive.

\emph{Model corpus.} The model side is a local gateway exposing open-weight models
through an OpenAI-compatible interface. We first probe every model the gateway
advertises with a trivial request and record whether it answers, is rate limited, or
is unavailable, so that only models that genuinely respond enter the study. From the
responding set we select \NModels{} models chosen for diversity of family and size,
listed in Table~\ref{tab:models}. For each model and subject we send the subject's
prompt verbatim, with a fixed system instruction asking for a single complete code
solution, and store the returned code with the model, the resolved model name, the
latency, and the token counts. Generation is repeated so that each model contributes
more than one sample per subject, and the whole run is resumable and paced to respect
the gateway's rate limits. No output is edited or synthesised. The gateway throttles
models independently and unpredictably, so coverage is filled by repeated resumable
passes rather than forced, and a missing model-subject cell is reported as missing.

\emph{Pipeline.} Listing~\ref{lst:pipeline} states the collection and analysis
pipeline in full. It is a deterministic sequence of small steps, each writing an
artifact to disk, and the fail-closed checker of Appendix~\ref{app:repro} re-derives
every reported number from those artifacts. Adding a subject is a data change, a
single specification entry rather than a code change, so the study extends to a new
task without modifying the pipeline.

\noindent\begin{minipage}{\linewidth}
\begin{lstlisting}[caption={Collection and analysis pipeline.},label={lst:pipeline}]
for subject in specification:   # one entry per task
  human[subject] = stackoverflow_api(subject.query)
  for model in usable_models:   # probed WORKS models
    for r in range(repeats):
      code = gateway.chat(model, subject.prompt)
      save(code, provenance=(model, latency, tokens))

for sample in human + model_samples:
  checks = [regex(c) for c in subject.checks]  # secure/insecure
  style  = style_metrics(sample.code)          # size, structure
  sec    = secure_frac(checks) - insecure_frac(checks)
  write_features(sample, checks, style, sec)

attribution = cross_val(RandomForest, features, provenance)
prevalence  = adopt_rate(checks) by (subject, source)
verify(every_macro == recomputed_from(features))  # fail closed
\end{lstlisting}
\end{minipage}

\section{Feature Extraction}
\label{sec:features}

Every sample, human or model, is reduced to two families of features by
deterministic functions of the stored code, so that the reduction is reproducible
and interpretable.

\emph{Security-pattern checks.} Each subject carries a list of checks, and each
check is a regular expression tagged as secure or insecure. A check fires when its
pattern is present in the code. For OAuth with PKCE, for example, the secure checks
include the presence of a code challenge and verifier following PKCE~\cite{rfc7636}, an anti-forgery state
parameter, and validation of the redirect target, while an insecure check fires on a
hardcoded client secret. For password hashing, a secure check fires on a strong key
derivation function such as bcrypt, scrypt, argon2, or pbkdf2, and an insecure check
fires on a bare md5 or sha1. From the checks we derive, for each sample, the fraction
of secure patterns adopted, the fraction of insecure patterns present, and a single
security score,
\begin{equation}
\text{sec} = \frac{\#\text{secure adopted}}{\#\text{secure defined}}
           - \frac{\#\text{insecure present}}{\#\text{insecure defined}},
\end{equation}
which lies in $[-1, 1]$ and is higher for code that adopts the secure practices and
avoids the insecure ones.

\emph{Style metrics.} Independently of the task, we compute language-agnostic style
metrics that capture the shape of the code, its length in characters and lines, the
comment density, the number of import statements, the blank-line ratio, the average
line length, and indicators for the presence of functions and of error handling.
These are the features that code stylometry uses, in a lightweight and interpretable
form, and they carry most of the provenance signal of \S\ref{sec:provenance}.

\section{Dataset}
\label{sec:dataset}

The collected corpus contains \NSamples{} real code samples, \NHuman{} from Stack
Overflow and \NLLM{} generated by \NModels{} open-weight models, spread across
\NSubjects{} subjects. The subjects reduce to \NFamilies{} task families expressed
across \NLanguages{} languages, namely Python, JavaScript, Go, and a markup task in
HTML, which is what makes the cross-language analysis of \S\ref{sec:crosslang} possible. Table~\ref{tab:subjects} lists the subjects with the number
of security and style checks each carries and the count of human and model samples
retained. Table~\ref{tab:models} lists the models with their family, approximate
size, and sample count. The model set spans several families and a wide size range,
from small general models to very large mixture-of-experts systems, and includes both
code-specialised and general models. Because the gateway throttles each backend
independently, coverage is not uniform, and we report the true per-model counts
rather than forcing an equal cell count.

\begin{table}[tbp]
\centering
\caption{The \NSubjects{} security-sensitive subjects. SO and LLM are the retained
human and model sample counts.}
\label{tab:subjects}
\small
\begin{tabular}{@{}p{3.5cm}llrr@{}}
\toprule
Subject & Lang & Chk & SO & LLM \\
\midrule
OAuth 2.0 Authorization Code flow with PKCE & python & 6 & 7 & 18 \\
JWT verification for API authentication & python & 5 & 8 & 17 \\
Password hashing for user storage & python & 4 & 6 & 15 \\
Database query with user input & python & 3 & 8 & 14 \\
Handling a user file upload & python & 4 & 8 & 15 \\
CORS configuration for an API & python & 3 & 8 & 15 \\
Responsive product landing page & html & 6 & 0 & 13 \\
Session cookie setup & python & 4 & 3 & 14 \\
OAuth 2.0 Authorization Code flow with PKCE (Node) & javascript & 6 & 0 & 14 \\
JWT verification for API authentication (Node) & javascript & 5 & 4 & 14 \\
Password hashing for user storage (Node) & javascript & 4 & 0 & 14 \\
Database query with user input (Node) & javascript & 3 & 0 & 14 \\
CORS configuration for an API (Node) & javascript & 3 & 8 & 14 \\
Handling a user file upload (Node) & javascript & 4 & 0 & 14 \\
Session cookie setup (Node) & javascript & 4 & 1 & 14 \\
Password hashing for user storage (Go) & go & 4 & 0 & 12 \\
JWT verification for API authentication (Go) & go & 5 & 8 & 12 \\
Database query with user input (Go) & go & 3 & 0 & 12 \\
CORS configuration for an API (Go) & go & 3 & 8 & 12 \\
Handling a user file upload (Go) & go & 4 & 0 & 12 \\
Session cookie setup (Go) & go & 4 & 0 & 12 \\
OAuth 2.0 Authorization Code flow with PKCE (Go) & go & 6 & 0 & 12 \\
Running a shell command with user input & python & 3 & 7 & 12 \\
Rendering user input in an HTML page & python & 3 & 4 & 12 \\
Fetching a URL provided by the user & python & 3 & 2 & 12 \\
Loading an API key or secret in an app & python & 2 & 5 & 12 \\
Serving a file by name from a directory & python & 2 & 1 & 12 \\
OAuth 2.0 login with Spring Security (Java) & java & 6 & 8 & 12 \\
OAuth 2.0 login with Django (Python) & python & 6 & 2 & 12 \\
OAuth 2.0 login with Express and Passport (Node) & javascript & 5 & 3 & 12 \\
Database query with user input in Spring (Java) & java & 3 & 8 & 12 \\
\bottomrule
\end{tabular}

\end{table}

\begin{table}[tbp]
\centering
\caption{The \NModels{} open-weight models used, with family, approximate size, and number of retained samples.}
\label{tab:models}
\scriptsize
\begin{tabular}{@{}lllr@{}}
\toprule
Model & Family & Size & $n$ \\
\midrule
\texttt{\scriptsize codestral} & Mistral & 22B & 68 \\
\texttt{\scriptsize mistral-large-3-675b} & Mistral & 675B & 30 \\
\texttt{\scriptsize mistral-small-4-119b} & Mistral & 119B & 62 \\
\texttt{\scriptsize nemotron-3-nano-30b} & Nemotron & 30B & 62 \\
\texttt{\scriptsize nemotron-3-ultra-550b} & Nemotron & 550B & 60 \\
\texttt{\scriptsize north-mini-code} & North & mini & 62 \\
\texttt{\scriptsize qwen3-coder-30b} & Qwen & 30B & 1 \\
\texttt{\scriptsize qwen3.5-397b} & Qwen & 397B & 4 \\
\texttt{\scriptsize tencent-hy3} & Tencent & MoE & 62 \\
\bottomrule
\end{tabular}

\end{table}

\section{Provenance Attribution}
\label{sec:provenance}

\emph{Human versus model.} We first ask whether a sample's provenance can be
recovered from its features alone. We train a random forest~\cite{scikit} on the style metrics of
\S\ref{sec:features} together with the two aggregate security fractions, and evaluate
it with stratified five-fold cross-validation against a majority-class baseline. The
classifier separates human from model code with \AttrAcc{} percent accuracy against a
\AttrBase{} percent baseline, at an F1 of \AttrFone{} for the model class. Provenance is
therefore highly recoverable from lightweight, interpretable features, without any
learned representation of the code.

\begin{figure}[tbp]
\centering
\includegraphics[width=0.92\columnwidth]{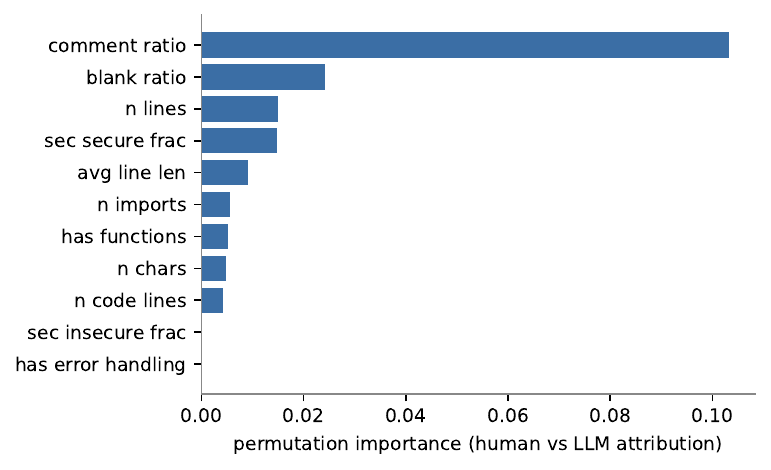}
\caption{Permutation importance of each feature for human-versus-model attribution.
Code shape, not security content, carries most of the provenance signal.}
\label{fig:importance}
\end{figure}

Figure~\ref{fig:importance} reports the permutation importance of each feature for
this task. The signal is dominated by the shape of the code rather than its security
content. The strongest discriminators are the raw size of the sample, the presence of
functions, and the number of imports, which reflects a consistent difference in how
much scaffolding each source produces for the same request. The security fractions
contribute, but they are secondary to style, so the human-versus-model boundary is
primarily stylistic rather than security-based.

\emph{Which model.} We then restrict to the model samples and ask a harder
question, which specific model wrote a given sample. A \ModelAttrWays{}-way random
forest over the same features reaches \ModelAttrAcc{} percent accuracy against a
\ModelAttrBase{} percent majority baseline. Individual models therefore leave a
recoverable fingerprint in their code, well above chance, though the boundary between
models is far less sharp than the boundary between human and machine. Model
attribution from code alone is a real but bounded capability, consistent with the
view that models within a shared training regime converge on overlapping styles.

\begin{figure}[tbp]
\centering
\includegraphics[width=\columnwidth]{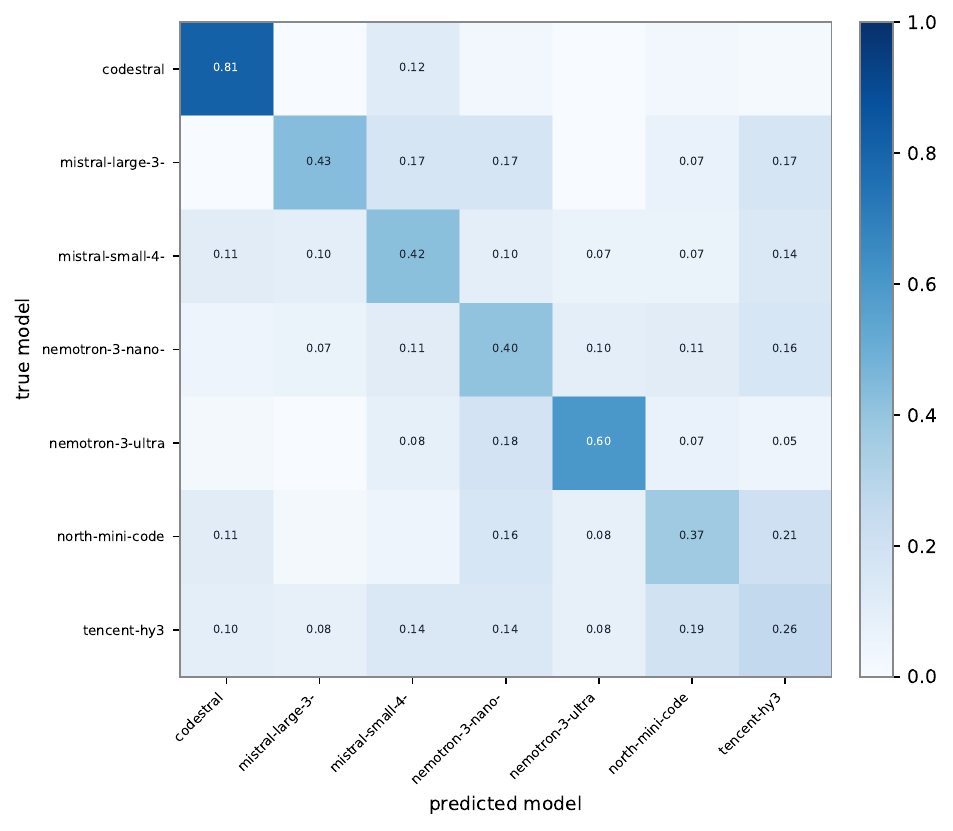}
\caption{Row-normalised confusion matrix of the model-attribution classifier.
Same-family models are confused with each other more than across families.}
\label{fig:modelconf}
\end{figure}

Figure~\ref{fig:modelconf} shows the row-normalised confusion matrix of the model
classifier. The structure is informative. Models from the same family are confused
with each other more than with models from other families, and the models that
produce the most elaborate scaffolding are the most cleanly separated. The confusions
follow family and size rather than being uniform noise, which is what a genuine
per-model fingerprint predicts.

\section{Security Divergence}
\label{sec:security}

\begin{figure*}[tp]
\centering
\includegraphics[width=\textwidth]{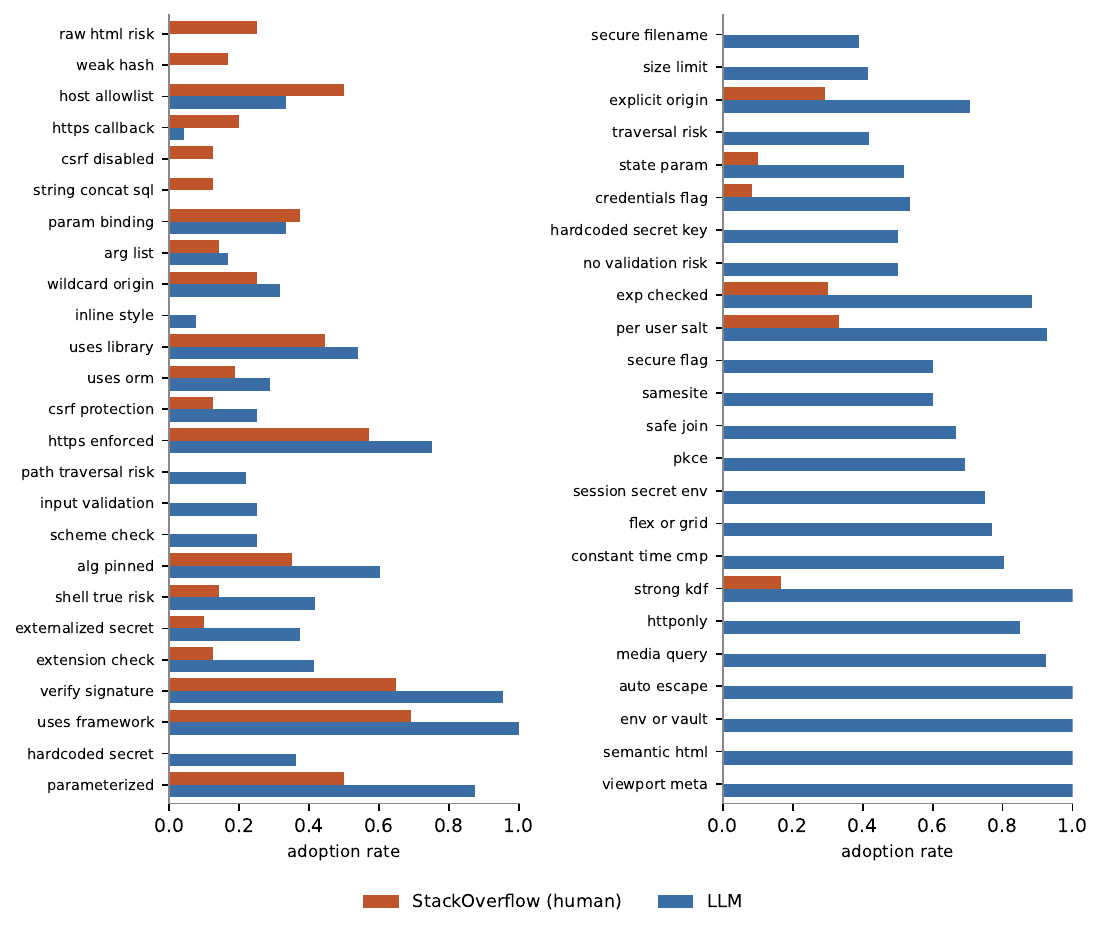}
\caption{Per-pattern adoption, human Stack Overflow versus model, aggregated over all
subjects and sorted by the model-minus-human gap. Secure defensive patterns are
adopted far more often by the models, while a small number of insecure patterns,
notably hardcoded secrets, are more common in model code. Patterns that neither source
adopts are omitted.}
\label{fig:prevalence}
\end{figure*}

The second question is whether the two sources adopt the same security patterns for
the same task. Table~\ref{tab:prevalence} reports, for each detector, the adoption
rate in the human corpus and in the model corpus, aggregated across subjects and
weighted by sample count. Figure~\ref{fig:prevalence} plots the same contrast.

\begin{table}[tbp]
\centering
\caption{Adoption rate of each security and style pattern in the human Stack Overflow
corpus and in the model corpus, aggregated across subjects. Higher is more frequent.}
\label{tab:prevalence}
\scriptsize
\begin{tabular}{@{}lrr@{}}
\toprule
Pattern & SO adopt & LLM adopt \\
\midrule
\code{alg none risk} & 0.00 & 0.02 \\
\code{alg pinned} & 0.35 & 0.60 \\
\code{alt text} & -- & 0.00 \\
\code{arg list} & 0.14 & 0.17 \\
\code{auto escape} & 0.00 & 1.00 \\
\code{constant time cmp} & 0.00 & 0.80 \\
\code{credentials flag} & 0.08 & 0.54 \\
\code{csp} & 0.00 & 0.00 \\
\code{csrf disabled} & 0.12 & 0.00 \\
\code{csrf protection} & 0.12 & 0.25 \\
\code{debug on} & 0.00 & 0.00 \\
\code{env or vault} & 0.00 & 1.00 \\
\code{exp checked} & 0.30 & 0.88 \\
\code{explicit origin} & 0.29 & 0.71 \\
\code{extension check} & 0.12 & 0.41 \\
\code{externalized secret} & 0.10 & 0.38 \\
\code{flex or grid} & -- & 0.77 \\
\code{hardcoded key} & 0.00 & 0.00 \\
\code{hardcoded secret} & 0.00 & 0.36 \\
\code{hardcoded secret key} & 0.00 & 0.50 \\
\code{host allowlist} & 0.50 & 0.33 \\
\code{httponly} & 0.00 & 0.85 \\
\code{https callback} & 0.20 & 0.04 \\
\code{https enforced} & 0.57 & 0.75 \\
\code{inline style} & -- & 0.08 \\
\code{input validation} & 0.00 & 0.25 \\
\code{media query} & -- & 0.92 \\
\code{no validation risk} & 0.00 & 0.50 \\
\code{param binding} & 0.38 & 0.33 \\
\code{parameterized} & 0.50 & 0.88 \\
\code{path traversal risk} & 0.00 & 0.22 \\
\code{per user salt} & 0.33 & 0.93 \\
\code{pkce} & 0.00 & 0.69 \\
\code{raw html risk} & 0.25 & 0.00 \\
\code{redirect validation} & 0.00 & 0.00 \\
\code{safe join} & 0.00 & 0.67 \\
\code{samesite} & 0.00 & 0.60 \\
\code{scheme check} & 0.00 & 0.25 \\
\code{secure filename} & 0.00 & 0.39 \\
\code{secure flag} & 0.00 & 0.60 \\
\code{semantic html} & -- & 1.00 \\
\code{session secret env} & 0.00 & 0.75 \\
\code{shell true risk} & 0.14 & 0.42 \\
\code{size limit} & 0.00 & 0.41 \\
\code{state param} & 0.10 & 0.52 \\
\code{string concat sql} & 0.12 & 0.00 \\
\code{strong kdf} & 0.17 & 1.00 \\
\code{traversal risk} & 0.00 & 0.42 \\
\code{uses framework} & 0.69 & 1.00 \\
\code{uses library} & 0.44 & 0.54 \\
\code{uses orm} & 0.19 & 0.29 \\
\code{verify signature} & 0.65 & 0.95 \\
\code{viewport meta} & -- & 1.00 \\
\code{weak hash} & 0.17 & 0.00 \\
\code{wildcard origin} & 0.25 & 0.32 \\
\bottomrule
\end{tabular}

\end{table}

\begin{figure*}[tp]
\centering
\includegraphics[width=0.92\textwidth]{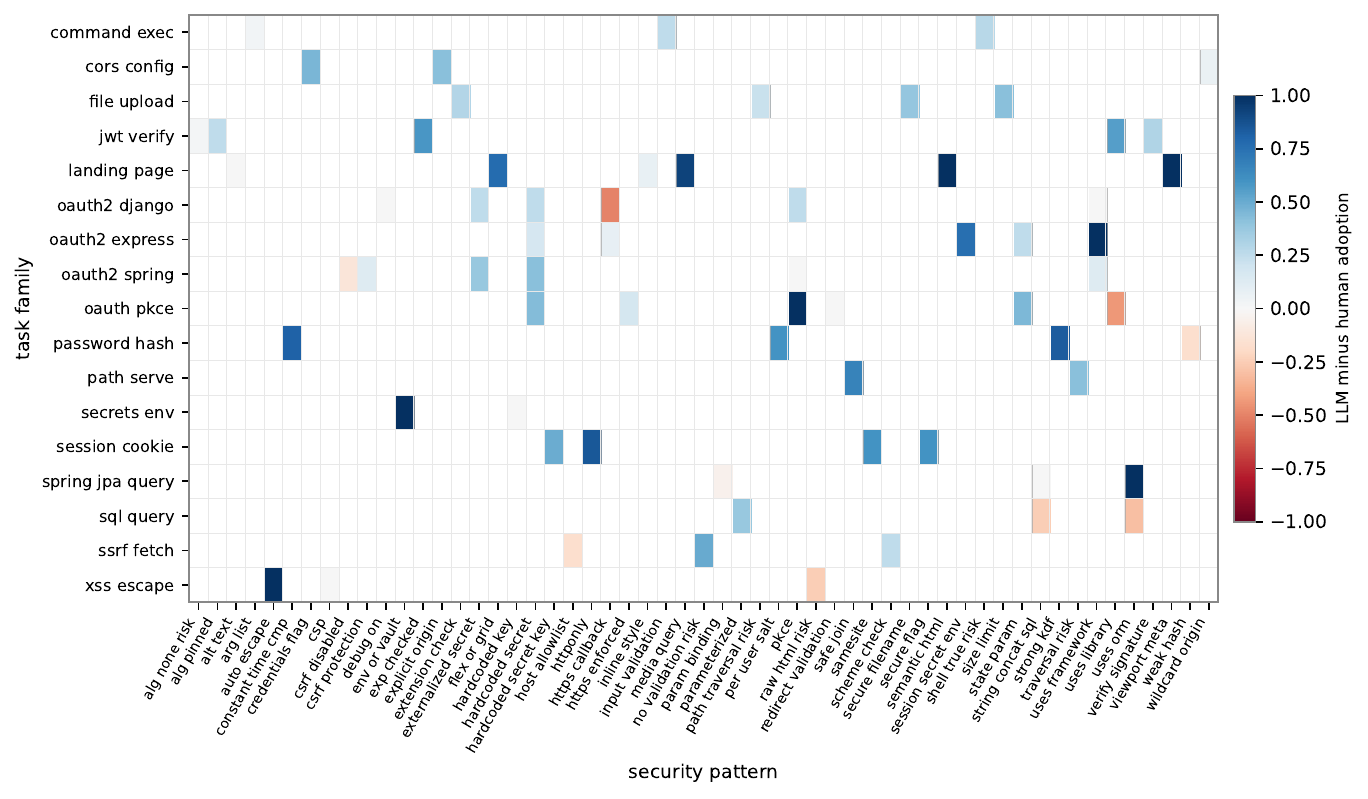}
\caption{Signed difference in pattern adoption, model minus human, per task family and
pattern (language variants collapsed to families). Blue marks patterns the models
adopt more, red marks patterns the human corpus adopts more, and a blank cell means
the pattern does not apply to that family.}
\label{fig:prevheat}
\end{figure*}

Figure~\ref{fig:prevheat} gives a per-family, per-pattern view of the same contrast
as the signed difference in adoption between the models and the human corpus, so the
blue cells are patterns the models adopt more and the red cells are patterns the human
corpus adopts more. The figure makes the structure of the divergence visible at a
glance, a broad band of blue on the defensive patterns and a small number of red cells
on the patterns where the human corpus leads.

\emph{Models adopt defensive patterns more often.} The clearest result is that the
model corpus adopts the secure patterns at a substantially higher rate than the human
corpus. The aggregate security score, the fraction of secure patterns adopted minus
the fraction of insecure patterns present, is \SecLLM{} for the models against
\SecHuman{} for the human corpus. The gap is largest on exactly the patterns that
modern secure-coding guidance emphasises. The models reach for proof-key-for-code
exchange in the OAuth task, a strong key-derivation function for password storage, a
parameterized query for database access, signature verification and algorithm pinning
and expiry checking for tokens, and the secure, http-only, and same-site flags for
session cookies, all at markedly higher rates than the human answers, which frequently
omit them. This is consistent with the models having absorbed the more recent
consensus on these tasks, while the highly-voted human answers include material old
enough to predate it.

\emph{Where the human corpus carries insecure legacy patterns.} The same table
shows the human corpus retaining insecure habits that the models have largely shed.
Human answers concatenate user input into SQL, use bare or fast hash functions for
passwords, and configure a wildcard cross-origin policy more often than the models do.
These are the classic copy-and-paste hazards, and their lower rate in model code is
the mirror image of the previous finding.

\emph{The exception, hardcoded secrets.} The models are not uniformly safer. On one
family of patterns they are worse. Model code introduces a hardcoded application
secret key and a hardcoded client secret more often than the human corpus, typically
as an inline placeholder value in an otherwise complete program. This is a real and
recurring hazard. A placeholder secret that ships unchanged is a live vulnerability,
and it is a pattern the human answers, which more often leave configuration to the
reader, produce less. The security picture is therefore not that models are safe and
humans are not, but that the two sources fail in different places, the human corpus on
legacy defensive omissions and the models on inlined secrets and completeness that
invites unedited reuse.

\begin{figure}[tbp]
\centering
\includegraphics[width=\columnwidth]{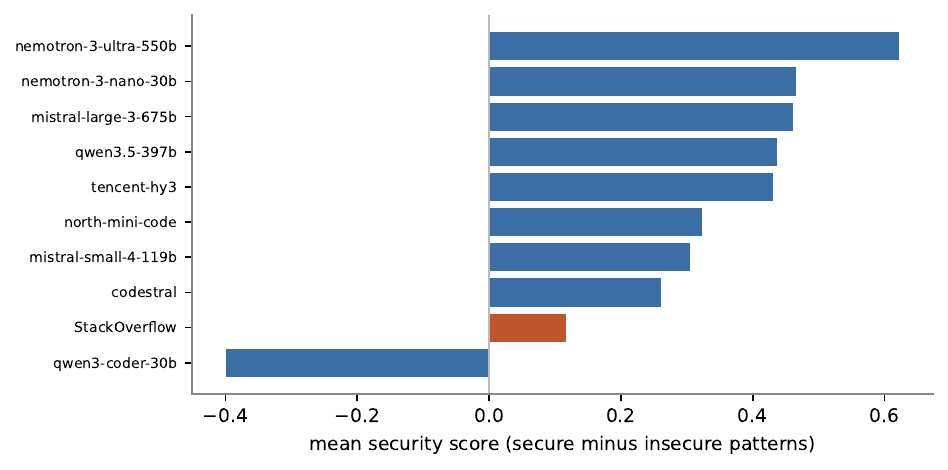}
\caption{Mean security score by origin, the human corpus alongside each model. The
human corpus is lowest and the models spread above it.}
\label{fig:secbymodel}
\end{figure}

\emph{Security score by origin.} Figure~\ref{fig:secbymodel} breaks the security
score down by origin, with the human corpus alongside each model. The human corpus
sits at the bottom, and the models spread above it, so the aggregate gap is not driven
by a single model but is a property of the model corpus as a whole. The spread among
models is itself informative, and \S\ref{sec:security} discusses which models sit
highest.

\section{Style Divergence}
\label{sec:style}

The provenance result of \S\ref{sec:provenance} is driven by style, and
Table~\ref{tab:style} makes the style contrast concrete. For the same task, model
code is far larger than the human answer, \LLMChars{} characters on average against
\HumanChars{}, roughly a threefold difference. The models wrap their solution in
functions, add error handling, and import more supporting modules, where the human
answers are terser fragments that assume a surrounding context. Error handling is
present in \LLMErrH{} percent of model samples against \HumanErrH{} percent of human
samples. Human answers use longer individual lines, a density that comes from writing
a minimal snippet rather than a complete program. These differences are the substance
of the attribution result, and they also explain the completeness hazard of
\S\ref{sec:security}, since a longer, self-contained, runnable program is exactly the
kind of artifact a developer is tempted to paste and ship with its placeholder secret
intact.

\begin{table}[tbp]
\centering
\caption{Mean style features by source. Model code is longer, more structured, and
more defensively wrapped than the human answers for the same task.}
\label{tab:style}
\small
\begin{tabular}{@{}lrr@{}}
\toprule
Feature & StackOverflow & LLM \\
\midrule
n code lines & 14.368 & 50.148 \\
comment ratio & 0.054 & 0.17 \\
n imports & 0.444 & 1.555 \\
has error handling & 0.085 & 0.314 \\
avg line len & 51.364 & 35.321 \\
sec score & 0.117 & 0.401 \\
\bottomrule
\end{tabular}

\end{table}

\begin{figure}[tbp]
\centering
\includegraphics[width=0.82\columnwidth]{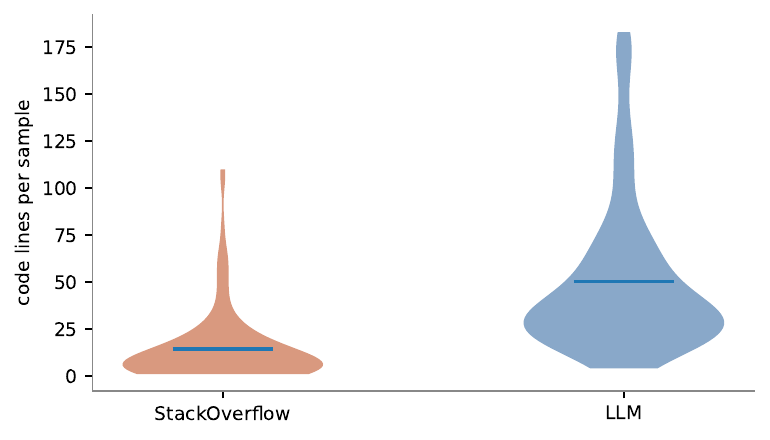}
\caption{Distribution of code length in lines, by source. The human and model
distributions barely overlap.}
\label{fig:length}
\end{figure}

\begin{figure}[tbp]
\centering
\includegraphics[width=0.9\columnwidth]{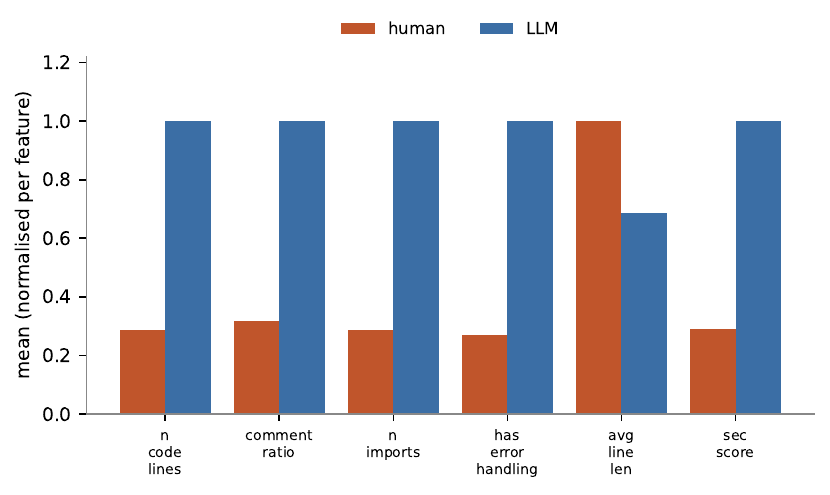}
\caption{Style features by source, each normalised across the two sources. Model code
is larger and more structured, human code is denser per line.}
\label{fig:style}
\end{figure}

Figure~\ref{fig:style} normalises each style feature across the two sources for a
direct visual comparison, and shows the same pattern, models higher on size,
structure, and error handling, humans higher on line density.
Figure~\ref{fig:length} shows the distribution of code length behind the mean, and
the two distributions barely overlap, which is why length alone is a strong provenance
signal.

\section{Per-Subject Analysis}
\label{sec:persubject}

\begin{figure*}[tp]
\centering
\includegraphics[width=0.74\textwidth]{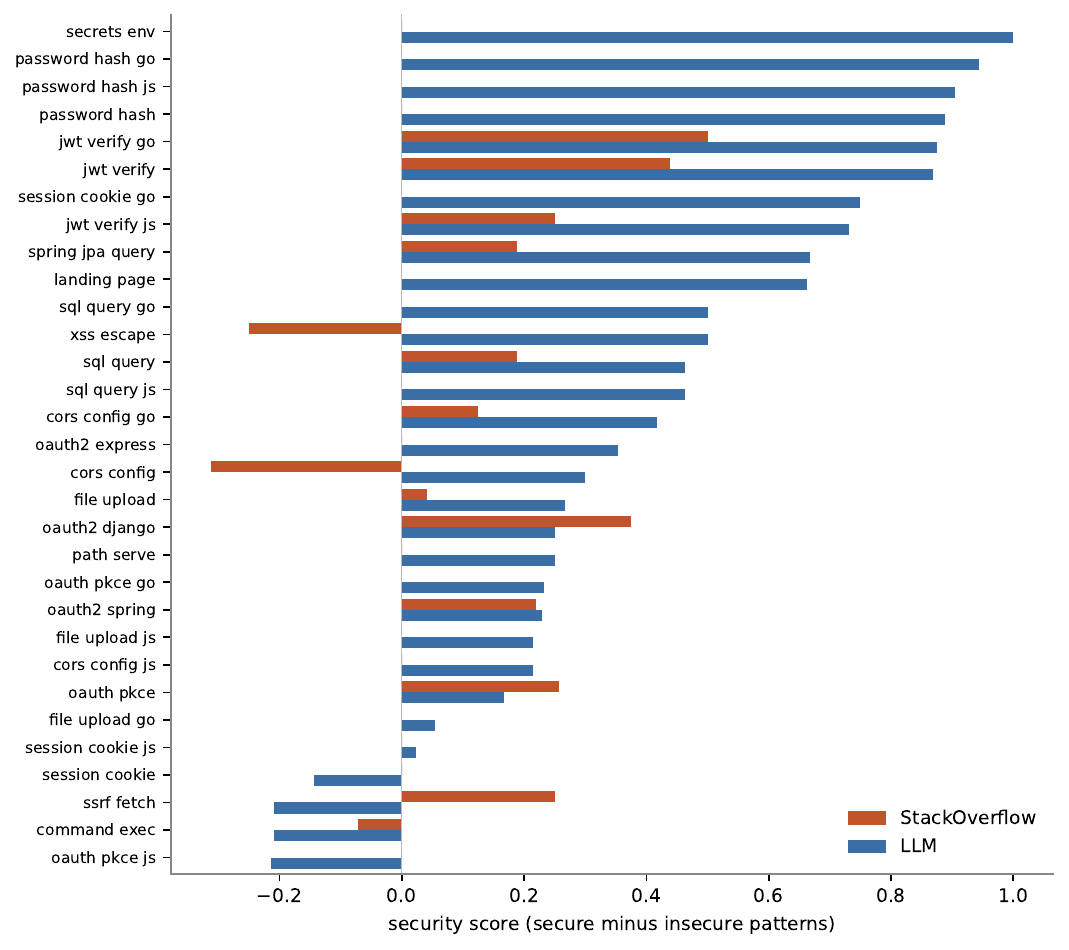}
\caption{Security score by subject, human Stack Overflow versus model, sorted by the
model score. The models are far safer on password hashing and token verification,
roughly even or worse on OAuth, SSRF, and session cookies, where their
hardcoded-secret and missing-defense habits cost them.}
\label{fig:subjectsec}
\end{figure*}

The aggregate result of \S\ref{sec:security} hides real per-subject structure. The
models are not uniformly safer than the human corpus. On some tasks they are far
safer, on others they are worse, and the reasons differ by task.
Figure~\ref{fig:subjectsec} gives the security score of each source on each subject,
and this section walks through the subjects that carry the clearest lesson, showing a
real human answer and a real model generation for each.

\subsection{Password Hashing}

Password hashing is the subject with the sharpest divide. The human corpus scores at
the floor, and the models near the ceiling. The reason is visible in the code. The
representative human answer, Listing~\ref{lst:password_hash_h}, reaches for a fast,
general-purpose hash, the exact pattern that decades of guidance warn against for
passwords, and it is a highly-voted answer precisely because it is old. The
representative model answer, Listing~\ref{lst:password_hash_l}, uses a purpose-built
password key-derivation function with a per-user salt and a constant-time verify.
This is the cleanest case of the models carrying the current consensus while the human
archive preserves the practice that consensus replaced.

\noindent\begin{minipage}{\linewidth}
\begin{lstlisting}[caption={Human SO answer 39660610 (score 9), password hash},label={lst:password_hash_h}]
>>> from passlib.hash import pbkdf2_sha256
>>> pbkdf2_sha256.hash("password", rounds=200000, salt=b'spamhameggs')
'$pbkdf2-sha256$200000$c3BhbWhhbWVnZ3M$WL9OLVcb3f7HqHeNT./kCJeunydL...
\end{lstlisting}
\end{minipage}

\noindent\begin{minipage}{\linewidth}
\begin{lstlisting}[caption={LLM codestral, password hash},label={lst:password_hash_l}]
import bcrypt

def hash_password(password: str) -> str:
    """
    Securely hash a user password for storage in a database.

    Args:
        password: The plain text password to be hashed.

    Returns:
        The hashed password as a string.
    """
    # Generate a salt and hash the password
    salt = bcrypt.gensalt()
    hashed_password = bcrypt.hashpw(password.encode('utf-8'), salt)
    return hashed_password.decode('utf-8')

def verify_password(stored_hash: str, provided_password: str) -> bool:
    """
    Verify a user's password against the stored hash.
# ...
\end{lstlisting}
\end{minipage}

\subsection{Token Verification}

The JWT subject repeats the pattern. The human answers frequently decode a token
without pinning the accepted algorithms or checking expiry, the ingredients of the
classic algorithm-confusion and replay problems. The model answers, one of which is
Listing~\ref{lst:jwt_verify_l}, verify the signature, pin the algorithm list, and
check expiry by default. The human example, Listing~\ref{lst:jwt_verify_h}, shows the
terser style that omits these steps because it answers a narrower question.

\noindent\begin{minipage}{\linewidth}
\begin{lstlisting}[caption={Human SO answer 43145184 (score 29), jwt verify},label={lst:jwt_verify_h}]
#!/usr/bin/env python

import jwt
from cryptography.x509 import load_pem_x509_certificate
from cryptography.hazmat.backends import default_backend

PEMSTART = "-----BEGIN CERTIFICATE-----\n"
PEMEND = "\n-----END CERTIFICATE-----\n"

mspubkey = "The value from the x5c property"
IDTOKEN = "the id_token to be validated"
tenant_id = "your tenant id"

cert_str = PEMSTART + mspubkey + PEMEND
cert_obj = load_pem_x509_certificate(cert_str, default_backend())
public_key = cert_obj.public_key()

decoded = jwt.decode(IDTOKEN, public_key, algorithms=['RS256'], aud...
if decoded:
    print "Decoded!"
# ...
\end{lstlisting}
\end{minipage}

\noindent\begin{minipage}{\linewidth}
\begin{lstlisting}[caption={LLM mistral-large-3-675b, jwt verify},label={lst:jwt_verify_l}]
import jwt
from jwt.exceptions import InvalidTokenError
from functools import wraps
from flask import request, jsonify  # Assuming Flask; adjust for yo...

# Configuration - replace with your actual values
SECRET_KEY = "your-secret-key"  # Use environment variables in prod...
ALGORITHM = "HS256"  # Algorithm used to sign the token

def verify_jwt_token(f):
    """Decorator to verify JWT token in API requests."""
    @wraps(f)
    def wrapper(*args, **kwargs):
        auth_header = request.headers.get("Authorization")
        if not auth_header or not auth_header.startswith("Bearer "):
            return jsonify({"error": "Authorization token is missin...

        token = auth_header.split(" ")[1]
        try:
            payload = jwt.decode(token, SECRET_KEY, algorithms=[ALG...
# ...
\end{lstlisting}
\end{minipage}

\subsection{Cross-Origin Policy}

CORS is the subject where the human corpus scores below zero, its insecure patterns
outnumbering its secure ones. The common human answer opens the API to any origin
with a wildcard, the fastest way to make a cross-origin call succeed and a standing
security hazard. The model answers more often restrict the allowed origins to an
explicit list. Listings~\ref{lst:cors_config_h} and~\ref{lst:cors_config_l} show the
contrast.

\noindent\begin{minipage}{\linewidth}
\begin{lstlisting}[caption={Human SO answer 27280939 (score 375), cors config},label={lst:cors_config_h}]
// The following property can be used to configure cross-origin res...
// in the HTTP nodes.
// See https://github.com/troygoode/node-cors#configuration-options...
// details on its contents. The following is a basic permissive set...
httpNodeCors: {
  origin: "*",
  methods: "GET,PUT,POST,DELETE"
},
\end{lstlisting}
\end{minipage}

\noindent\begin{minipage}{\linewidth}
\begin{lstlisting}[caption={LLM north-mini-code, cors config},label={lst:cors_config_l}]
from flask import Flask, jsonify
from flask_cors import CORS

# Initialize the Flask application
app = Flask(__name__)

# Configure CORS for the application
# This allows the frontend on a different origin to make requests.
# For production, restrict 'origins' to specific domains (e.g., "ht...
CORS(app, resources={r"/*": {"origins": "*"}})

@app.route('/api/data')
def get_data():
    """
    Example endpoint returning JSON data.
    """
    return jsonify({"message": "CORS is working!", "status": "succe...

if __name__ == '__main__':
    # Run the development server
# ...
\end{lstlisting}
\end{minipage}

\subsection{OAuth and Session Cookies, Where the Models Slip}

The two subjects where the models do not lead are the ones that expose their own
failure mode. On OAuth the models reliably add the proof-key-for-code-exchange steps
that the human answers omit, yet their overall score is dragged down because they also
inline a client secret as a placeholder in the same file. On session cookies the same
thing happens with the application secret key. The models produce a complete,
runnable configuration, and completeness here means filling the secret slot with a
literal value, which the terser human answers leave to the reader.
Listings~\ref{lst:oauth_pkce_h} and~\ref{lst:oauth_pkce_l} show the OAuth case, where
the model code is longer and more correct on the protocol yet carries the inlined
secret that the audit of \S\ref{sec:security} flags.

\noindent\begin{minipage}{\linewidth}
\begin{lstlisting}[caption={Human SO answer 29629536 (score 46), oauth pkce},label={lst:oauth_pkce_h}]
from flask import url_for, current_app, redirect, request
from rauth import OAuth2Service

import json, urllib2

class OAuthSignIn(object):
    providers = None

    def __init__(self, provider_name):
        self.provider_name = provider_name
        credentials = current_app.config['OAUTH_CREDENTIALS'][provi...
        self.consumer_id = credentials['id']
        self.consumer_secret = credentials['secret']

    def authorize(self):
        pass

    def callback(self):
        pass
# ...
\end{lstlisting}
\end{minipage}

\noindent\begin{minipage}{\linewidth}
\begin{lstlisting}[caption={LLM mistral-small-4-119b, oauth pkce},label={lst:oauth_pkce_l}]
import secrets
import base64
import hashlib
import urllib.parse
import requests
from typing import Dict, Tuple

class OAuthPKCE:
    """
    OAuth 2.0 Authorization Code Flow with PKCE implementation.
    Handles authorization request generation and token exchange.
    """

    def __init__(self, client_id: str, client_secret: str, redirect...
                 auth_url: str, token_url: str, scope: str = None):
        self.client_id = client_id
        self.client_secret = client_secret
        self.redirect_uri = redirect_uri
        self.auth_url = auth_url
        self.token_url = token_url
# ...
\end{lstlisting}
\end{minipage}

\subsection{Parameterized Queries}

The database subject sits between the extremes. The human corpus is split. Some
answers use a parameterized query, the safe form, while a substantial fraction build
the query by concatenating or formatting the user input directly into the SQL string,
the textbook injection hazard. The models almost always parameterize.
Listings~\ref{lst:sql_query_h} and~\ref{lst:sql_query_l} show a human and a model
answer. The human corpus also reaches for an object-relational mapper more often than
the models, which hand-write the parameterized query, a difference of idiom that does
not change the security outcome but does change the style.

\noindent\begin{minipage}{\linewidth}
\begin{lstlisting}[caption={Human SO answer 25029149 (score 21), sql query},label={lst:sql_query_h}]
from django.shortcuts import render
from django.shortcuts import HttpResponse
from .models import Person
from django.core.exceptions import *

def index(request):
    return render(request, 'form.html')

def search(request):
    if request.method == 'POST':
        search_id = request.POST.get('textfield', None)
        try:
            user = Person.objects.get(name = search_id)
            #do something with user
            html = ("<H1>%s</H1>", user)
            return HttpResponse(html)
        except Person.DoesNotExist:
            return HttpResponse("no such user")
    else:
        return render(request, 'form.html')
\end{lstlisting}
\end{minipage}

\noindent\begin{minipage}{\linewidth}
\begin{lstlisting}[caption={LLM mistral-small-4-119b, sql query},label={lst:sql_query_l}]
import sqlite3
from typing import Optional

def get_user_by_username(username: str) -> Optional[dict]:
    """
    Look up a user in a SQLite database by username.

    Args:
        username: The username to search for.

    Returns:
        A dictionary representing the user if found, None otherwise.
    """
    conn = None
    try:
        conn = sqlite3.connect('users.db')  # Replace with your act...
        cursor = conn.cursor()

        cursor.execute("SELECT * FROM users WHERE username = ?", (u...
        user = cursor.fetchone()
# ...
\end{lstlisting}
\end{minipage}

\subsection{File Upload}

File upload is a subject where neither source scores well, and the reason is the
number of separate defenses the task needs. A safe upload validates the file type,
sanitizes the filename against path traversal, and limits the size, and few samples
from either source do all three. The models do somewhat more of the validation than
the human answers, as Listings~\ref{lst:file_upload_h} and~\ref{lst:file_upload_l}
show, but both sources leave gaps, which makes upload a subject where provenance is a
weaker guide to safety than hashing or tokens.

\noindent\begin{minipage}{\linewidth}
\begin{lstlisting}[caption={Human SO answer 62357616 (score 7), file upload},label={lst:file_upload_h}]
#!/usr/bin/env python
# -*- coding: utf-8 -*-
from flask import Flask
from flask import request
from flask import render_template
import os

app = Flask(__name__)


@app.route("/", methods=['POST', 'GET'])
def index():
    if request.method == "POST":
        f = request.files['audio_data']
        with open('audio.wav', 'wb') as audio:
            f.save(audio)
        print('file uploaded successfully')

        return render_template('index.html', request="POST")
    else:
# ...
\end{lstlisting}
\end{minipage}

\noindent\begin{minipage}{\linewidth}
\begin{lstlisting}[caption={LLM mistral-large-3-675b, file upload},label={lst:file_upload_l}]
from fastapi import FastAPI, UploadFile, File, HTTPException
from fastapi.responses import JSONResponse
import os
from pathlib import Path

app = FastAPI()

# Ensure upload directory exists
UPLOAD_DIR = Path("uploads")
UPLOAD_DIR.mkdir(exist_ok=True)

@app.post("/upload/")
async def upload_file(file: UploadFile = File(...)):
    """
    Endpoint to upload a file. The file will be saved to the 'uploa...
    Returns the saved file path on success.
    """
    try:
        # Create a secure file path
        file_path = UPLOAD_DIR / file.filename
# ...
\end{lstlisting}
\end{minipage}

\subsection{Injection and the Limits of the Trend}

The injection-style subjects, added to probe tasks where the danger is subtler than a
missing library call, complicate the clean picture of the earlier subjects and are the
most interesting for it. On some of them the models keep their lead. For rendering user
input into a page, the models reach for an auto-escaping template almost always, while
the human answers more often build the HTML by hand and leave the escaping out, so the
cross-site-scripting hazard is markedly lower in model code. For loading a secret, the
models read it from an environment variable or a secret store, where the human answers
more often inline a literal.

On other injection subjects the models lose their lead or fall behind, and the reason
is again their objective. Asked to run a shell command with a user-supplied argument,
the models invoke the shell directly, the convenient and dangerous form, more often than
the human answers do, because it is the shortest complete solution. Asked to fetch a
user-supplied URL, the models frequently fetch it with no validation of the scheme or
the host, missing the server-side-request-forgery defense entirely, because nothing in
the prompt named it and the happy-path solution does not need it. These are the same
completeness hazard seen with secrets, in a more dangerous form. The model writes the
whole feature and omits the defense that a security-aware developer would add against a
threat the prompt did not mention. The lesson of the injection subjects is that the
model advantage holds where the safe choice is a well-known library and evaporates, or
reverses, where the safe choice is to anticipate an unmentioned threat.
Listings~\ref{lst:xss_escape_l} and~\ref{lst:ssrf_fetch_l} put the two directions side
by side, a model auto-escaping the cross-site-scripting task and a model fetching a
user URL with no server-side-request-forgery check in the same corpus.

\noindent\begin{minipage}{\linewidth}
\begin{lstlisting}[caption={LLM codestral, xss escape},label={lst:xss_escape_l}]
from flask import Flask, request, render_template_string

app = Flask(__name__)

# HTML template with a form to input a message and display it
HTML_TEMPLATE = """
<!DOCTYPE html>
<html>
<head>
    <title>Message Display</title>
</head>
<body>
    <h1>Enter a Message</h1>
    <form method="POST">
        <input type="text" name="message" placeholder="Enter your m...
        <button type="submit">Submit</button>
    </form>
    {% if message %}
        <h2>Your Message:</h2>
        <p>{{ message }}</p>
# ...
\end{lstlisting}
\end{minipage}

\noindent\begin{minipage}{\linewidth}
\begin{lstlisting}[caption={LLM tencent-hy3, ssrf fetch},label={lst:ssrf_fetch_l}]
import httpx
from fastapi import FastAPI, HTTPException, Query
from fastapi.responses import Response

app = FastAPI(title="URL Fetcher")

@app.get("/fetch")
async def fetch_url(url: str = Query(..., description="The URL to f...
    # Note: In production, validate/allowlist URLs to prevent SSRF ...
    try:
        async with httpx.AsyncClient(follow_redirects=True) as client:
            response = await client.get(url, timeout=10.0)
            response.raise_for_status()
            # Return the fetched content directly with its original...
            return Response(content=response.content, media_type=re...
    except httpx.HTTPError as e:
        raise HTTPException(status_code=502, detail=f"Error fetchin...

if __name__ == "__main__":
    import uvicorn
# ...
\end{lstlisting}
\end{minipage}

\noindent\begin{minipage}{\linewidth}
\begin{lstlisting}[caption={Human SO answer 52544846 (score 39), command exec},label={lst:command_exec_h}]
import subprocess


def start(executable_file):
    return subprocess.Popen(
        executable_file,
        stdin=subprocess.PIPE,
        stdout=subprocess.PIPE,
        stderr=subprocess.PIPE)


def read(process):
    return process.stdout.readline().decode("utf-8").strip()


def write(process, message):
    process.stdin.write(f"{message.strip()}\n".encode("utf-8"))
    process.stdin.flush()
# ...
\end{lstlisting}
\end{minipage}

\noindent\begin{minipage}{\linewidth}
\begin{lstlisting}[caption={LLM mistral-small-4-119b, command exec},label={lst:command_exec_l}]
import subprocess

def run_shell_command_with_file(filename: str) -> None:
    """Run a shell command with a user-provided filename."""
    # Example command: cat the file and count lines
    command = ["wc", "-l", filename]

    try:
        result = subprocess.run(
            command,
            check=True,
            text=True,
            stdout=subprocess.PIPE,
            stderr=subprocess.PIPE,
        )
        print(f"Command output:\n{result.stdout}")
    except subprocess.CalledProcessError as e:
        print(f"Command failed with error:\n{e.stderr}")
    except FileNotFoundError:
        print("Error: 'wc' command not found. Adjust command as nee...
# ...
\end{lstlisting}
\end{minipage}

\subsection{Summary of the Per-Subject Picture}

Across the subjects the lesson is consistent with the aggregate but more precise. The
models lead by a wide margin wherever the secure practice is a well-known library call
that the human archive predates, hashing and token verification most of all. They lose
their lead, and occasionally fall behind, wherever the secure practice is to leave a
value unset, because their objective pushes them to produce a complete program and a
complete program fills every slot. Provenance therefore predicts not just whether code
is likely safe but which specific hazard to look for, a library-omission hazard in
human code and a completeness hazard in model code.

\section{Framework-Level Comparison}
\label{sec:framework}

The subjects so far use a minimal library so that the security choice is isolated. A
developer, however, asks a framework-level question, for example how to set up OAuth
2.0 login in a Java Spring Boot application, or how to add social login to Django, or
how to wire up Passport in Express. We add four such subjects, phrased as the natural
question a developer would type, and send the identical prompt to every model. Because
every model answers the same question, we can place their implementations side by side
and compare which security practices each one adopts.

Table~\ref{tab:fwcompare} does this for the Spring Security OAuth subject. Each row is
one model's answer to the same prompt, and each column is a security or configuration
check, marked secure or insecure. The comparison makes the differences between models
concrete. All of the models produce a working Spring Security configuration, but they
diverge on the details that matter. Some enable proof-key-for-code exchange and read
the client secret from externalised configuration, while others inline the secret or
disable the cross-site-request-forgery protection that Spring enables by default, the
kind of convenience shortcut that is common in tutorials and dangerous in production.
The same question therefore yields materially different security postures depending on
which model answers it, and the table names exactly where.

\begin{table*}[t]
\centering
\caption{Framework comparison for the Spring Security OAuth subject. Each row is one
model's answer to the same prompt. A check mark means the pattern is present. Columns
marked (S) are secure practices and (I) are insecure ones.}
\label{tab:fwcompare}
\small
\setlength{\tabcolsep}{10pt}
\begin{tabular}{@{}lcccccc@{}}
\toprule
Model & fw & PKCE & CSRF+ & CSRF off & hard secret & env secret \\
 & (S) & (S) & (S) & (I) & (I) & (S) \\
\midrule
\texttt{\scriptsize codestral} & \checkmark &  &  &  & \checkmark & \checkmark \\
\texttt{\scriptsize mistral-small-4-119b} & \checkmark &  &  &  &  & \checkmark \\
\texttt{\scriptsize nemotron-3-nano-30b} & \checkmark &  & \checkmark &  & \checkmark &  \\
\texttt{\scriptsize nemotron-3-ultra-550b} & \checkmark &  &  &  &  &  \\
\texttt{\scriptsize north-mini-code} & \checkmark &  &  &  &  &  \\
\texttt{\scriptsize tencent-hy3} & \checkmark &  & \checkmark &  &  & \checkmark \\
\bottomrule
\end{tabular}

\end{table*}

The framework subjects also confirm the completeness hazard of \S\ref{sec:security} in
a realistic setting. The framework answers are longer and more complete than the
minimal-library answers, and the extra completeness again carries risk, since a model
that produces a full runnable Spring configuration is more likely to fill the client
secret with a literal placeholder than one that leaves the wiring to the reader. The
practical implication for a developer is direct. The convenience of a complete,
copy-ready framework answer is exactly what makes it worth re-checking for an inlined
secret and for a disabled default protection before it is used.

\section{Cross-Language Analysis}
\label{sec:crosslang}

The core tasks are defined in three programming languages, Python, JavaScript, and Go,
for the same underlying task, which lets us ask whether the two findings of the paper
are properties of a language or of the source. We treat each task as a family with a
variant per language, and compare within and across them. The security comparison uses
all three languages. The provenance-transfer experiment, which needs a dense paired
sample, uses the Python and JavaScript variants where coverage is deepest.

\emph{The security gap holds in every language.} The central security result does not
depend on the language. In each of the three the model corpus sits well above the
human corpus, a mean security score of \SecPyLLM{} against the human corpus in Python
and \SecJsLLM{} in JavaScript, with Go higher for both sources but the same gap between
them. The absolute level differs by ecosystem, Go scoring higher because its standard
library steers even the human answers toward safer defaults, but the direction and the
size of the human-to-model gap are consistent across the three. Table~\ref{tab:familylang}
gives the per-family breakdown, and the pattern from \S\ref{sec:security} repeats. The
models lead on the library-backed defensive tasks, hashing and token verification, and
slip on the tasks where the safe choice is to leave a secret unset. That the same
structure appears in three languages is strong evidence that it reflects how the two
sources approach a task rather than a quirk of one ecosystem.

\begin{table}[tbp]
\centering
\caption{Model security score per task family in each language. The divergence
structure of \S\ref{sec:security} repeats across languages.}
\label{tab:familylang}
\small
\begin{tabular}{@{}lrrrr@{}}
\toprule
Task family & Py sec & Py $n$ & JS sec & JS $n$ \\
\midrule
\code{command exec} & -0.21 & 12 & -- & -- \\
\code{cors config} & +0.30 & 15 & +0.21 & 14 \\
\code{cors config go} & -- & -- & -- & -- \\
\code{file upload} & +0.27 & 15 & +0.21 & 14 \\
\code{file upload go} & -- & -- & -- & -- \\
\code{jwt verify} & +0.87 & 17 & +0.73 & 14 \\
\code{jwt verify go} & -- & -- & -- & -- \\
\code{landing page} & -- & -- & -- & -- \\
\code{oauth2 django} & +0.25 & 12 & -- & -- \\
\code{oauth2 express} & -- & -- & +0.35 & 12 \\
\code{oauth2 spring} & -- & -- & -- & -- \\
\code{oauth pkce} & +0.17 & 18 & -0.21 & 14 \\
\code{oauth pkce go} & -- & -- & -- & -- \\
\code{password hash} & +0.89 & 15 & +0.91 & 14 \\
\code{password hash go} & -- & -- & -- & -- \\
\code{path serve} & +0.25 & 12 & -- & -- \\
\code{secrets env} & +1.00 & 12 & -- & -- \\
\code{session cookie} & -0.14 & 14 & +0.02 & 14 \\
\code{session cookie go} & -- & -- & -- & -- \\
\code{spring jpa query} & -- & -- & -- & -- \\
\code{sql query} & +0.46 & 14 & +0.46 & 14 \\
\code{sql query go} & -- & -- & -- & -- \\
\code{ssrf fetch} & -0.21 & 12 & -- & -- \\
\code{xss escape} & +0.50 & 12 & -- & -- \\
\bottomrule
\end{tabular}

\end{table}

\begin{figure*}[tp]
\centering
\includegraphics[width=\textwidth]{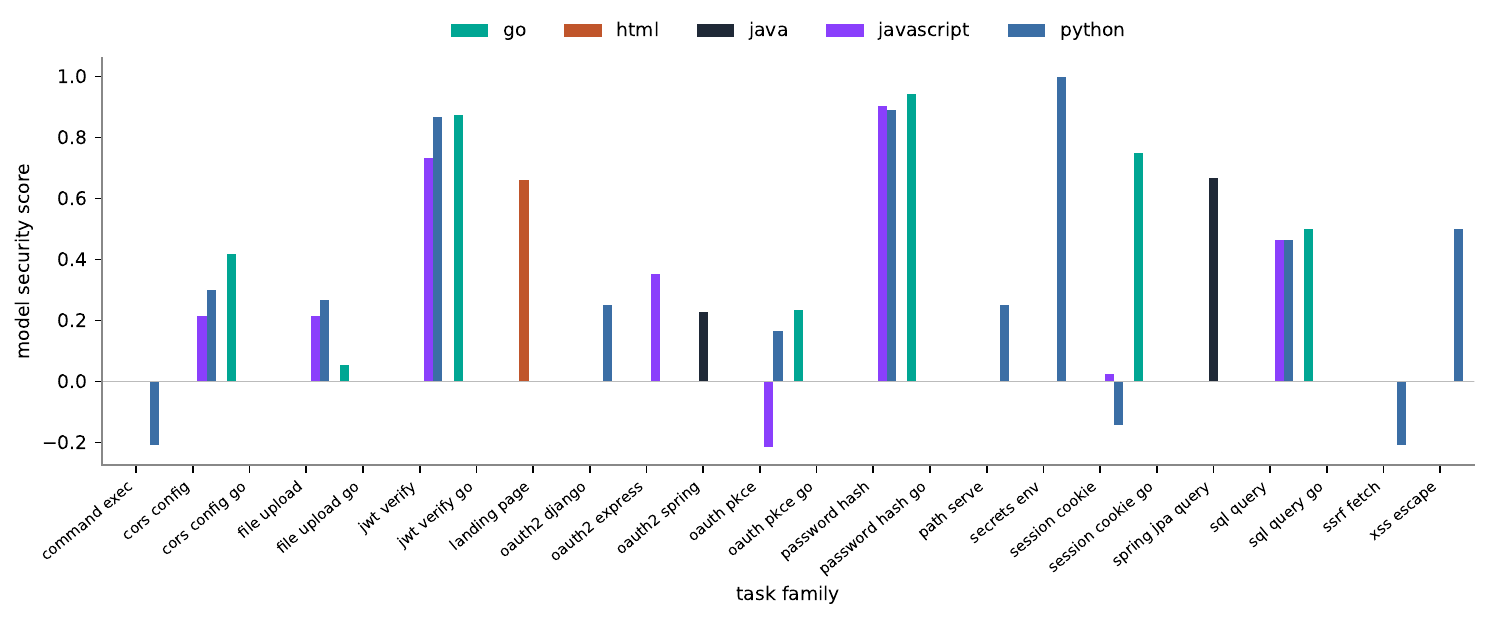}
\caption{Model security score per task family, one bar per language. The languages
rise and fall together, so the security divergence is a property of the source rather
than of one ecosystem.}
\label{fig:familylang}
\end{figure*}

Figure~\ref{fig:familylang} plots the same per-family scores as grouped bars, one bar
per language, and the visual impression is of two profiles that rise and fall together.
Where the models are strong in one language, on hashing and token verification, they
are strong in the other, and where they slip, on the tasks that reward leaving a
secret unset, they slip in both. The correlation of the per-family scores across
languages is the quantitative form of this observation, and it is high, which is the
core evidence that the security divergence is a property of the source rather than of
a single language ecosystem.

\emph{Style varies by language.} Table~\ref{tab:langstyle} reports the model style
per language, and the reason the provenance boundary does not transfer becomes
concrete. The models write a different amount of scaffolding in each language, driven
by the conventions of the ecosystem, so the very features that carry the provenance
signal, length and import count and structure, take different baseline values from one
language to the next. A classifier that learned the model style of one language is
therefore reading a different scale when it sees another, which is the mechanism behind
the transfer failure below.

\begin{table}[tbp]
\centering
\caption{Model code style per language. The amount of scaffolding differs by language,
which is why the provenance boundary does not transfer freely.}
\label{tab:langstyle}
\small
\begin{tabular}{@{}lrrrr@{}}
\toprule
Language & Lines & Imports & Comment & Err.\ handling \\
\midrule
go & 62.88 & 1.0 & 0.17 & 0.01 \\
html & 136.77 & 0.0 & 0.06 & 0.0 \\
java & 35.62 & 3.88 & 0.13 & 0.0 \\
javascript & 49.64 & 0.35 & 0.22 & 0.43 \\
python & 40.2 & 2.35 & 0.15 & 0.45 \\
\bottomrule
\end{tabular}

\end{table}

\emph{Provenance signal is partly language-specific.} The attribution result behaves
differently across languages, and the asymmetry is informative. A human-versus-
model classifier trained on one language and tested on the other does not transfer
symmetrically. Trained on JavaScript and tested on Python it reaches \XLangJsPy{}
percent, close to its within-language accuracy, but trained on Python and tested on
JavaScript it falls to \XLangPyJs{} percent, near chance. The provenance signal
therefore has a language-general component, the gross size and structure difference
that survives the transfer in one direction, and a language-specific component that
does not. The asymmetry suggests that the model style learned on Python is narrower
than the style learned on JavaScript, so a classifier calibrated on the richer
JavaScript signal still recognises the Python one, while the reverse does not hold.
This is a caution for anyone deploying a provenance classifier. It must be calibrated
on the language it will see, because the human-versus-model boundary does not sit in
the same place in every language.

\emph{The same task in three languages.} The concreteness of the cross-language
result is easiest to see in the code. Listings~\ref{lst:password_hash_l},
\ref{lst:password_hash_js_l}, and~\ref{lst:password_hash_go_l} show a model solving the
password-hashing task in Python, JavaScript, and Go. The three differ in every surface
detail, the imports, the idioms, the shape, which is why a provenance classifier does
not transfer freely between them. All three, however, use the same class of defense, a
purpose-built password hash with a salt and a safe verify, which is why the security
score is the same across the languages. The listings show both parts of the result,
that the style is language-specific and the security posture is not.

\noindent\begin{minipage}{\linewidth}
\begin{lstlisting}[caption={LLM nemotron-3-nano-30b, password hash js},label={lst:password_hash_js_l}]
// password-utils.js
// Securely hash and verify passwords using the industry standard b...
// Install with: npm install bcrypt

const bcrypt = require('bcrypt');

/**
 * Generate a salt and hash a plaintext password.
 * @param {string} password - The user's plaintext password.
 * @returns {Promise<string>} A promise that resolves to the hashed...
 */
async function hashPassword(password) {
  // 12 rounds is a good default; adjust based on your performance/...
  const saltRounds = 12;
  const hashed = await bcrypt.hash(password, saltRounds);
  return hashed;
}

/**
 * Verify a plaintext password against a stored hash.
// ...
\end{lstlisting}
\end{minipage}

\noindent\begin{minipage}{\linewidth}
\begin{lstlisting}[caption={LLM nemotron-3-ultra-550b, password hash go},label={lst:password_hash_go_l}]
package main

import (
    "fmt"
    "log"

    "golang.org/x/crypto/bcrypt"
)

// HashPassword generates a bcrypt hash of the plaintext password.
// The cost parameter (12) is a good default balancing security and...
func HashPassword(password string) (string, error) {
    bytes, err := bcrypt.GenerateFromPassword([]byte(password), 12)
    return string(bytes), err
}

// CheckPasswordHash compares a bcrypt hash with a plaintext password.
// Returns nil if they match, or an error (typically bcrypt.ErrMism...
func CheckPasswordHash(password, hash string) error {
    return bcrypt.CompareHashAndPassword([]byte(hash), []byte(passw...
// ...
\end{lstlisting}
\end{minipage}

\emph{Summary.} The languages agree on the substantive result and disagree on
the incidental one. The security divergence between human and model code is a property
of the source and appears in both languages, while the exact decision boundary that
separates human from model code is partly language-specific and does not transfer
freely. A study in a single language would have reported the first and silently
assumed the second.

\section{Case Study, Vulnerability Repair}
\label{sec:repair}

The generation results of \S\ref{sec:security} measure what a model writes when asked
for a task from scratch. A second, more practical question is what a model does when
it is handed insecure code and asked to fix it. This is the common real workflow of a
developer pasting a flagged snippet into an assistant, and it is a different capability
from clean-slate generation. We test it directly.

\emph{Setup.} We assemble \NSeeds{} short programs, each containing one well-known
vulnerability drawn from \NCWE{} distinct Common Weakness Enumeration classes, among
them a weak password hash (CWE-327), SQL injection (CWE-89), OS command injection
(CWE-78), a disabled signature check (CWE-347), a wildcard cross-origin policy
(CWE-942), a hardcoded credential (CWE-798), reflected cross-site scripting (CWE-79),
path traversal (CWE-22), a non-cryptographic random source (CWE-330), server-side
request forgery (CWE-918), unsafe deserialization (CWE-502), and an open redirect
(CWE-601). The seeds span four languages, Python, JavaScript, Go, and Java. Each seed carries two regular expressions, one
for the insecure pattern that a fix must remove and one for the secure pattern a
correct fix must introduce. We send each seed to every model with a neutral
instruction to fix the security vulnerability and return the corrected code, and we
score the result deterministically. A repair counts as successful only when the
insecure pattern is gone \emph{and} the expected secure pattern is present, so a model
that deletes the dangerous call without adding the correct defense is not credited.
This yields \NRepairs{} scored repairs.

\emph{Models repair most but not all vulnerabilities.} The overall repair success rate
is \FixRate{} percent. The models are competent at the task, and Table~\ref{tab:repair}
shows the per-model breakdown, with the strongest model reaching \BestRepairRate{}
percent. The spread across models is real and follows capability, with the larger
models repairing more reliably than the smaller ones, which matches the intuition that
repair, like generation, rewards a model that knows the idiomatic secure replacement.

\begin{table}[tbp]
\centering
\caption{Vulnerability-repair success rate by model. A repair succeeds only when the
insecure pattern is removed and the correct secure pattern is added.}
\label{tab:repair}
\small
\begin{tabular}{@{}lrrr@{}}
\toprule
Model & Repairs & Fixed & Rate \\
\midrule
\texttt{\scriptsize nemotron-3-ultra-550b} & 20 & 18 & 0.90 \\
\texttt{\scriptsize north-mini-code} & 19 & 17 & 0.90 \\
\texttt{\scriptsize tencent-hy3} & 16 & 14 & 0.88 \\
\texttt{\scriptsize mistral-large-3-675b} & 14 & 12 & 0.86 \\
\texttt{\scriptsize nemotron-3-nano-30b} & 21 & 15 & 0.71 \\
\texttt{\scriptsize codestral} & 20 & 14 & 0.70 \\
\texttt{\scriptsize mistral-small-4-119b} & 21 & 11 & 0.52 \\
\bottomrule
\end{tabular}

\end{table}

\emph{The partial-fix hazard.} The most useful result of the case study is the failure
mode. In \RemoveNoAdd{} percent of all repairs the model removed the insecure pattern
but did not add the secure one. This is not a cosmetic miss. A repair that deletes an
md5 call but stores the password in plain text, or that removes a wildcard origin but
sets no origin at all, produces code that no longer matches the insecure signature yet
is not secure. A regex or signature-based scanner would mark such a repair as clean.
The partial fix is the repair-time analogue of the completeness hazard of
\S\ref{sec:security}, and it is a concrete caution against trusting an assistant's fix
without re-checking that the defense it was supposed to add is actually there.

\begin{figure*}[tp]
\centering
\includegraphics[width=0.86\textwidth]{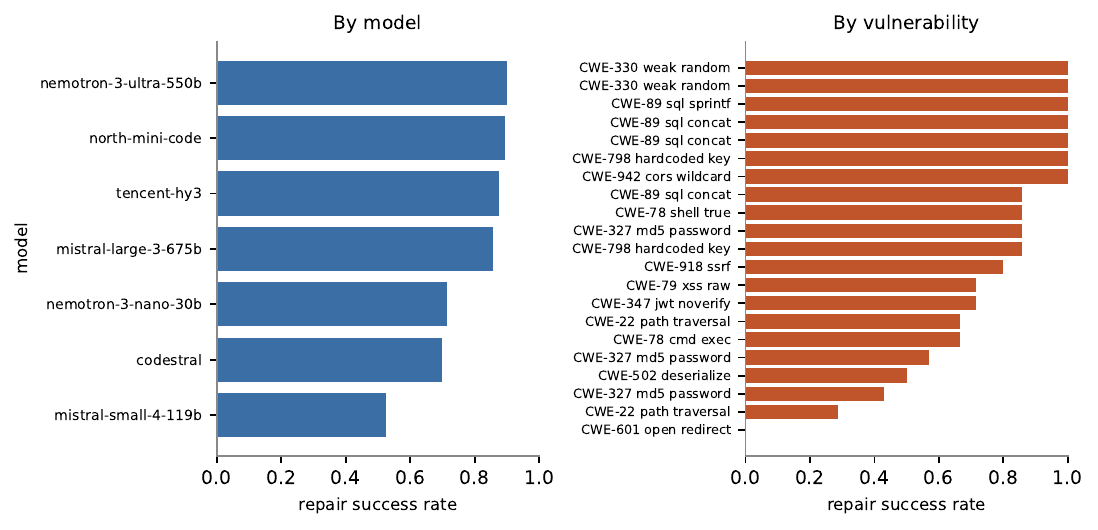}
\caption{Vulnerability-repair success rate by model (left) and by weakness class
(right). Library-substitution fixes succeed most often, and fixes that require adding
a missing check succeed least often.}
\label{fig:repair}
\end{figure*}

\emph{Some vulnerability classes are harder than others.}
Table~\ref{tab:repaircwe} and Figure~\ref{fig:repair} give the success rate per
vulnerability. The classes with a single canonical library fix, a parameterized query,
a strong password hash, or a cryptographic random source, are repaired most reliably,
often at or near a perfect rate, because the correct replacement is unambiguous and the
models know it. The classes that require adding a check the original code did not have,
validating a redirect target, a fetched URL, or a file path, are repaired least
reliably, the open redirect and the path traversal being the hardest in our set. The
reason is the same one that makes these classes written insecurely in the first place.
The safe form is an addition the model must think to make, not a substitution it can
pattern-match, and that is where both generation and repair are weakest. The pattern
holds across all four languages, with the ranking of easy and hard classes preserved,
so the effect is a property of the vulnerability class rather than of one language.

\begin{table}[tbp]
\centering
\caption{Vulnerability-repair success rate by weakness class, across all models.}
\label{tab:repaircwe}
\small
\begin{tabular}{@{}lllrr@{}}
\toprule
CWE & Vulnerability & Lang & $n$ & Rate \\
\midrule
CWE-601 & \code{open redirect} & py & 3 & 0.00 \\
CWE-22 & \code{path traversal} & py & 7 & 0.29 \\
CWE-327 & \code{md5 password} & ja & 7 & 0.43 \\
CWE-502 & \code{deserialize} & py & 4 & 0.50 \\
CWE-327 & \code{md5 password} & ja & 7 & 0.57 \\
CWE-78 & \code{cmd exec} & ja & 6 & 0.67 \\
CWE-22 & \code{path traversal} & ja & 6 & 0.67 \\
CWE-347 & \code{jwt noverify} & py & 7 & 0.71 \\
CWE-79 & \code{xss raw} & py & 7 & 0.71 \\
CWE-918 & \code{ssrf} & py & 5 & 0.80 \\
CWE-798 & \code{hardcoded key} & py & 7 & 0.86 \\
CWE-327 & \code{md5 password} & py & 7 & 0.86 \\
CWE-78 & \code{shell true} & py & 7 & 0.86 \\
CWE-89 & \code{sql concat} & ja & 7 & 0.86 \\
CWE-942 & \code{cors wildcard} & py & 7 & 1.00 \\
CWE-798 & \code{hardcoded key} & ja & 4 & 1.00 \\
CWE-89 & \code{sql concat} & ja & 7 & 1.00 \\
CWE-89 & \code{sql concat} & py & 7 & 1.00 \\
CWE-89 & \code{sql sprintf} & go & 7 & 1.00 \\
CWE-330 & \code{weak random} & ja & 5 & 1.00 \\
CWE-330 & \code{weak random} & py & 7 & 1.00 \\
\bottomrule
\end{tabular}

\end{table}

\section{Discussion}
\emph{Provenance is a stylistic boundary.} The attribution result of
\S\ref{sec:provenance} is strong, and its explanation matters as much as its accuracy.
The human-versus-model boundary is recovered mainly from the shape of the code, its
size, its structure, and its scaffolding, and only secondarily from its security
content. A reviewer or an audit tool can therefore flag likely machine-generated code
cheaply, from surface features alone, and the same features that make the code
distinguishable are the ones that make it feel finished. This connects the provenance
question to the security question. Machine code is recognisable because it is
complete, and its completeness is exactly what makes its embedded flaws dangerous to
paste unedited.

\emph{The security comparison is against a human baseline, not a gold standard.}
The finding that model code adopts secure patterns more often than the human corpus
must be read carefully. It is a comparison between two real sources a developer might
copy from, not a claim that model code is secure in absolute terms. The prior
literature, which compares model output against a correct reference, finds that
assistants still emit vulnerable code at a meaningful
rate~\cite{pearce2022asleep,perry2023users}. Our result is compatible with that. Against
the highly-voted but often dated human answers on Stack Overflow, the models have
shed many legacy omissions, yet they introduce their own failure mode. Both readings
hold, and the accurate summary is that the two sources fail differently rather than
that one is safe.

\emph{Why the sources fail where they do.} The pattern of divergence has a plausible
cause. The human corpus is an archive. A highly-voted answer accumulates its score
over years, and much of it predates the current defaults for hashing, token handling,
and cross-origin policy, which is why the human corpus under-adopts exactly those
defensive patterns. The models are trained on a snapshot that includes the more recent
guidance, so they reproduce the current consensus on those tasks. The models' own
failure, the inlined placeholder secret, follows from their objective. They are asked
for a single complete program, so they fill every slot, including the ones a human
answer would leave to the reader, and a filled secret slot is a hazard the human
fragment never creates.

\emph{Implications for review.} The two results combine into concrete guidance. Code
that a lightweight classifier flags as machine-generated should be routed to the
checks that model code most often fails, which the divergence profile of
\S\ref{sec:security} names, foremost the scan for hardcoded secrets and placeholder
configuration. Code identified as copied from a human source should instead be routed
to the legacy-omission checks, the hashing, token, and cross-origin defaults that the
human corpus most often misses. Provenance is thus not an end in itself but a router
that sends each artifact to the review it most needs.

\emph{Deploying a provenance classifier.} The cross-language result of
\S\ref{sec:crosslang} is a direct warning for anyone who would deploy the attribution
classifier as a tool. The boundary between human and model code does not sit in the
same place in every language, and a classifier calibrated on one language can perform
below chance on another, producing confident but incorrect labels. A deployable
detector must therefore be calibrated per language, and its confidence must be
discounted on any language it was not trained on. The interpretable features we use
make this tractable, because the small feature set can be re-fit on a modest
per-language sample rather than requiring a full retraining of a deep detector. It is
therefore more accurate to treat provenance attribution as a language-scoped
capability than as a universal one.

\emph{The moving-target caveat.} Both of our results are snapshots. The human corpus
ages in one direction, its highly-voted answers drifting further from current practice
as time passes, which will tend to widen the security gap we measure. The models move
in the other direction, retrained on newer data and tuned by their providers, which
may narrow the gap or shift the failure mode. The value of the released pipeline is
that it makes the measurement repeatable. The same specification re-run in a year will
report where each source has moved, and the fail-closed checker guarantees that the
new numbers are the ones the new data supports.

\section{Limitations and Threats to Validity}
\label{sec:limitations-lexical}
\emph{Pattern detectors are lexical.} The security checks are regular expressions,
not a semantic analysis. They detect the presence of a pattern, for example a call to
a strong key-derivation function, but they do not prove the pattern is used correctly.
A sample can adopt bcrypt and still store the result badly. The detectors are a
transparent, reproducible first cut, not a verifier, and the paper's claims are
therefore about pattern adoption rather than about proven security. We treat the
lexical nature of the detectors as the principal threat to construct validity and
avoid any claim that a high security score means a program is safe.

\emph{The human corpus is a snapshot of one site.} The human material is drawn from
Stack Overflow, retrieved through its public API with a fixed query per subject. A
different query or a different site would surface different code, and the age
distribution of highly-voted answers biases the corpus toward older practice, which is
part of what the security-divergence result measures rather than a nuisance to remove.
We record each answer's score and year so this bias is visible and auditable.

\emph{The model set is open-weight and gateway-limited.} The models are those a
public open-weight gateway makes available, and the widely-used closed commercial
assistants are not among them. The results therefore describe open-weight models,
which is a real and growing population, but not the specific commercial products the
earlier user studies examined. The gateway also throttles models unevenly, so
per-model sample counts differ, and a model with few samples contributes a noisier
security profile. We report true counts and weight aggregates by them.

\emph{Attribution is within a fixed pool.} The model-attribution result is a
closed-world classification among the models in the study. It does not address the
open-world question of recognising an unseen model, and its accuracy would fall as the
pool grows and model styles converge. We present it as evidence that a per-model
fingerprint exists at this scale, not as a deployable detector.

\emph{Prompt sensitivity.} Each subject uses a single fixed prompt. Model output is
sensitive to phrasing, and a prompt that explicitly asked for security would raise the
adoption of defensive patterns. Our prompts are deliberately neutral, phrased as an
ordinary developer request, so the measured adoption reflects the default behaviour of
each source rather than its best behaviour under prompting.

\section{Ethics and Responsible Use}
\label{sec:ethics}

The two capabilities this paper demonstrates, attributing code to a source and
profiling the security of that source, are dual use, and we state the intended and the
misuse cases plainly.

\emph{Intended use.} The provenance classifier is meant to route code to the review
it needs, as \S\ref{sec:security} describes, and to support consented uses such as a
team understanding the composition of its own codebase or an educator discussing the
difference between hand-written and generated solutions. The security-divergence
profile is meant to make secure-coding review more efficient by naming, per source,
the failures most worth looking for.

\emph{Misuse and why the risk is limited.} A provenance classifier could in principle
be used to penalise the use of an assistant, for example to detect and sanction
machine help where it is disallowed. We note two limits that reduce this risk. The
accuracy is bounded and, as \S\ref{sec:crosslang} shows, does not transfer across
languages, so a naive deployment would produce confident errors that make it unsuitable
as a sole basis for any consequential decision. And the signal is dominated by
surface style, which a user who wished to evade detection could alter trivially by
reformatting, so the classifier is not a robust adversarial detector and should not be
presented as one.

\emph{Data handling.} The human corpus is public Stack Overflow content retrieved
through the official API within its terms, and we store only the code block, its public
identifier, and its public score, no personal data. The model corpus is generated by us
through a gateway we are entitled to use, and no third-party private code is collected.
The released dataset therefore contains only public human answers and our own
generations.

\emph{No claim of absolute security.} The security scores are pattern-adoption
measures, not verification, as \S\ref{sec:limitations-lexical} states. We are careful
throughout not to certify any sample as secure, and we frame every result as a
comparison between sources rather than as an audit of any individual program.

\section{Conclusion}
We studied two sources of developer code side by side, the human answers of Stack
Overflow and the output of \NModels{} open-weight language models, on \NSubjects{}
security-sensitive tasks, using only open sources and a fully reproducible pipeline.
Two results stand out. Provenance is recoverable. A lightweight classifier separates
human from model code at \AttrAcc{} percent and attributes a sample to its specific
model at \ModelAttrAcc{} percent, and the boundary is stylistic before it is a matter
of security. Security diverges. Against the human corpus, the models adopt modern
defensive patterns far more often, with an aggregate security score of \SecLLM{}
against \SecHuman{}, yet they introduce their own hazard by inlining placeholder
secrets into otherwise complete programs. The two sources do not rank as safe and
unsafe. They fail in different places, and knowing which source a snippet came from
tells a reviewer which failure to look for. The pipeline is data driven and released,
so any new task is one specification entry away, and a fail-closed checker re-derives
every number here from the stored samples.

\appendices
\section{Pattern Detector Catalogue}
\label{app:checks}

Table~\ref{tab:checks} in Appendix~\ref{app:reftables} lists every security and style
detector used, its subject, its polarity (secure or insecure), and a short
description. Each detector is a regular expression evaluated against the sample text.
The catalogue is the complete, reproducible definition of what the security-divergence
analysis measures. A detector firing means its pattern is lexically present, which the
paper treats as adoption of the pattern, not as proof of its correct use.

\section{Per-Model Security Profile}
\label{app:modelsec}

Table~\ref{tab:modelsec} gives the mean security score and sample count for each
model and for the human corpus, the per-origin detail behind
Figure~\ref{fig:secbymodel}. The human corpus is included as a row so the models can
be read against it directly.

\begin{table}[!htb]
\centering
\caption{Mean security score and sample count by origin.}
\label{tab:modelsec}
\small
\setlength{\tabcolsep}{5pt}
\begin{tabular}{@{}lrr@{}}
\toprule
Origin & Sec. score & $n$ \\
\midrule
\code{nemotron-3-ultra-550b} & +0.62 & 60 \\
\code{nemotron-3-nano-30b} & +0.47 & 62 \\
\code{mistral-large-3-675b} & +0.46 & 30 \\
\code{qwen3.5-397b} & +0.44 & 4 \\
\code{tencent-hy3} & +0.43 & 62 \\
\code{north-mini-code} & +0.32 & 62 \\
\code{mistral-small-4-119b} & +0.30 & 62 \\
\code{codestral} & +0.26 & 68 \\
StackOverflow (human) & +0.12 & 117 \\
\code{qwen3-coder-30b} & -0.40 & 1 \\
\bottomrule
\end{tabular}

\end{table}

\section{Per-Model Security by Language}
\label{app:modellang}

Table~\ref{tab:modellang} gives each model's mean security score in each language, the per-model detail behind the cross-language result of \S\ref{sec:crosslang}. The scores are close across languages within a model, which is the model-level form of the finding that the security posture travels with the source rather than the language.

\begin{table}[!htb]
\centering
\caption{Mean security score per model in each language.}
\label{tab:modellang}
\small
\begin{tabular}{@{}lrrrrr@{}}
\toprule
Model & go & html & java & javascript & python \\
\midrule
\texttt{\scriptsize codestral} & +0.59 & +0.80 & +0.25 & +0.09 & +0.15 \\
\texttt{\scriptsize mistral-large-3-675b} & -- & +0.80 & -- & +0.40 & +0.47 \\
\texttt{\scriptsize mistral-small-4-119b} & +0.22 & +0.80 & +0.38 & +0.37 & +0.26 \\
\texttt{\scriptsize nemotron-3-nano-30b} & +0.78 & +0.10 & +0.38 & +0.36 & +0.41 \\
\texttt{\scriptsize nemotron-3-ultra-550b} & +0.62 & -- & +0.69 & +0.64 & +0.60 \\
\texttt{\scriptsize north-mini-code} & +0.51 & +0.60 & +0.50 & +0.14 & +0.28 \\
\texttt{\scriptsize qwen3-coder-30b} & -- & -- & -- & -- & -0.40 \\
\texttt{\scriptsize qwen3.5-397b} & -- & -- & -- & -- & +0.44 \\
\texttt{\scriptsize tencent-hy3} & +0.51 & +0.80 & +0.50 & +0.35 & +0.40 \\
\bottomrule
\end{tabular}

\end{table}

\section{Full Adoption Table}
\label{app:fullprev}

Table~\ref{tab:fullprev} in Appendix~\ref{app:reftables} gives the complete
per-subject, per-pattern adoption rate for the human corpus and the model corpus, the
detail behind the aggregate of \S\ref{sec:security} and the per-subject discussion of
\S\ref{sec:persubject}. H adopt and L adopt are the human and model adoption rates,
and the paired counts give the sample size behind each cell so that a rate resting on
few samples can be identified.

\section{Reproducibility}
\label{app:repro}

The study is reproducible end to end from public sources. The pipeline is a sequence
of small scripts, each writing an intermediate artifact to disk, and every number in
the paper is re-derived from those artifacts by a fail-closed checker.

\begin{enumerate}
  \item \code{probe\_fast.py} probes every model the gateway advertises and records,
  in \code{model\_availability.csv}, whether each answers, is rate limited, or is
  unavailable.
  \item \code{fetch\_stackoverflow.py} retrieves the human corpus for each subject
  through the public Stack Overflow API, storing each answer with its identifier,
  score, and year.
  \item \code{generate.py} sends each subject prompt to each usable model and stores
  the returned code with its provenance metadata. It is resumable and paced, and it
  is run as repeated passes so that models throttled during one pass are collected in
  a later one.
  \item \code{extract\_features.py} reduces every sample to the security and style
  features of \S\ref{sec:features}, writing \code{features.csv}.
  \item \code{benchmark.py} computes the attribution classifiers and the
  security-pattern prevalence, writing the analysis tables and \code{numbers.json}.
  \item \code{gen\_tables.py} and \code{gen\_figures.py} emit the paper's tables and
  figures, and \code{verify\_numbers.py} re-derives every macro from the data and
  fails if any disagrees.
\end{enumerate}

The subject specification, the model list, and every collected sample are released
with the code, so a third party can re-run the pipeline or extend it to a new subject
by adding a single specification entry. The classifiers use scikit-learn~\cite{scikit} with a fixed
random seed, so the reported accuracies are deterministic given the same samples.

\onecolumn
\section{Reference Tables}
\label{app:reftables}

The two reference tables below are the full detector catalogue and the complete
per-subject adoption data. They are set in a single column so they can run across
pages without truncation.

\begin{longtable}{@{}p{3.2cm}p{3.2cm}c p{7.0cm}@{}}
\caption{Complete detector catalogue. Polarity S marks a secure pattern and I an insecure one.}\label{tab:checks}\\
\toprule
Subject & Pattern & Pol. & Description \\
\midrule
\endfirsthead
\multicolumn{4}{c}{\footnotesize\itshape Table \thetable\ continued from previous page}\\
\toprule Subject & Pattern & Pol. & Description \\ \midrule
\endhead
\midrule \multicolumn{4}{r}{\footnotesize\itshape continued}\\
\endfoot
\bottomrule
\endlastfoot
\code{\scriptsize oauth pkce} & \code{\scriptsize pkce} & S & \scriptsize uses PKCE (code challenge/verifier) \\
\code{\scriptsize oauth pkce} & \code{\scriptsize state param} & S & \scriptsize uses anti-CSRF state parameter \\
\code{\scriptsize oauth pkce} & \code{\scriptsize redirect validation} & S & \scriptsize validates redirect uri \\
\code{\scriptsize oauth pkce} & \code{\scriptsize https enforced} & S & \scriptsize uses https endpoints \\
\code{\scriptsize oauth pkce} & \code{\scriptsize hardcoded secret} & I & \scriptsize hardcodes client secret \\
\code{\scriptsize oauth pkce} & \code{\scriptsize uses library} & S & \scriptsize uses an OAuth library vs hand-rolled \\
\code{\scriptsize jwt verify} & \code{\scriptsize verify signature} & S & \scriptsize verifies the signature \\
\code{\scriptsize jwt verify} & \code{\scriptsize alg pinned} & S & \scriptsize pins allowed algorithms \\
\code{\scriptsize jwt verify} & \code{\scriptsize alg none risk} & I & \scriptsize permits alg=none or verify=False \\
\code{\scriptsize jwt verify} & \code{\scriptsize exp checked} & S & \scriptsize checks expiry \\
\code{\scriptsize jwt verify} & \code{\scriptsize uses library} & S & \scriptsize uses a JWT library \\
\code{\scriptsize password hash} & \code{\scriptsize strong kdf} & S & \scriptsize uses bcrypt/scrypt/argon2/pbkdf2 \\
\code{\scriptsize password hash} & \code{\scriptsize weak hash} & I & \scriptsize uses md5/sha1/plain sha256 \\
\code{\scriptsize password hash} & \code{\scriptsize per user salt} & S & \scriptsize uses a salt \\
\code{\scriptsize password hash} & \code{\scriptsize constant time cmp} & S & \scriptsize constant-time comparison \\
\code{\scriptsize sql query} & \code{\scriptsize parameterized} & S & \scriptsize uses parameterized query \\
\code{\scriptsize sql query} & \code{\scriptsize string concat sql} & I & \scriptsize concatenates/format user input into SQL \\
\code{\scriptsize sql query} & \code{\scriptsize uses orm} & S & \scriptsize uses an ORM \\
\code{\scriptsize file upload} & \code{\scriptsize extension check} & S & \scriptsize validates file type/extension \\
\code{\scriptsize file upload} & \code{\scriptsize secure filename} & S & \scriptsize sanitizes the filename \\
\code{\scriptsize file upload} & \code{\scriptsize path traversal risk} & I & \scriptsize joins raw filename into a path \\
\code{\scriptsize file upload} & \code{\scriptsize size limit} & S & \scriptsize limits upload size \\
\code{\scriptsize cors config} & \code{\scriptsize wildcard origin} & I & \scriptsize allows any origin (*) \\
\code{\scriptsize cors config} & \code{\scriptsize explicit origin} & S & \scriptsize restricts to explicit origins \\
\code{\scriptsize cors config} & \code{\scriptsize credentials flag} & S & \scriptsize handles credentials explicitly \\
\code{\scriptsize landing page} & \code{\scriptsize viewport meta} & S & \scriptsize responsive viewport meta \\
\code{\scriptsize landing page} & \code{\scriptsize semantic html} & S & \scriptsize uses semantic elements \\
\code{\scriptsize landing page} & \code{\scriptsize alt text} & S & \scriptsize images have alt text \\
\code{\scriptsize landing page} & \code{\scriptsize media query} & S & \scriptsize uses responsive media queries \\
\code{\scriptsize landing page} & \code{\scriptsize flex or grid} & S & \scriptsize uses flexbox or grid \\
\code{\scriptsize landing page} & \code{\scriptsize inline style} & I & \scriptsize relies on inline styles \\
\code{\scriptsize session cookie} & \code{\scriptsize httponly} & S & \scriptsize sets HttpOnly \\
\code{\scriptsize session cookie} & \code{\scriptsize secure flag} & S & \scriptsize sets Secure \\
\code{\scriptsize session cookie} & \code{\scriptsize samesite} & S & \scriptsize sets SameSite \\
\code{\scriptsize session cookie} & \code{\scriptsize hardcoded secret key} & I & \scriptsize hardcodes the app secret key \\
\code{\scriptsize oauth pkce js} & \code{\scriptsize pkce} & S & \scriptsize uses PKCE (code challenge/verifier) \\
\code{\scriptsize oauth pkce js} & \code{\scriptsize state param} & S & \scriptsize uses anti-CSRF state parameter \\
\code{\scriptsize oauth pkce js} & \code{\scriptsize redirect validation} & S & \scriptsize validates redirect uri \\
\code{\scriptsize oauth pkce js} & \code{\scriptsize https enforced} & S & \scriptsize uses https endpoints \\
\code{\scriptsize oauth pkce js} & \code{\scriptsize hardcoded secret} & I & \scriptsize hardcodes client secret \\
\code{\scriptsize oauth pkce js} & \code{\scriptsize uses library} & S & \scriptsize uses an OAuth library \\
\code{\scriptsize jwt verify js} & \code{\scriptsize verify signature} & S & \scriptsize verifies the signature \\
\code{\scriptsize jwt verify js} & \code{\scriptsize alg pinned} & S & \scriptsize pins allowed algorithms \\
\code{\scriptsize jwt verify js} & \code{\scriptsize alg none risk} & I & \scriptsize permits alg none or decode without verify \\
\code{\scriptsize jwt verify js} & \code{\scriptsize exp checked} & S & \scriptsize checks expiry \\
\code{\scriptsize jwt verify js} & \code{\scriptsize uses library} & S & \scriptsize uses a JWT library \\
\code{\scriptsize password hash js} & \code{\scriptsize strong kdf} & S & \scriptsize uses bcrypt/scrypt/argon2 \\
\code{\scriptsize password hash js} & \code{\scriptsize weak hash} & I & \scriptsize uses md5/sha1 \\
\code{\scriptsize password hash js} & \code{\scriptsize per user salt} & S & \scriptsize uses a salt \\
\code{\scriptsize password hash js} & \code{\scriptsize constant time cmp} & S & \scriptsize safe compare / library verify \\
\code{\scriptsize sql query js} & \code{\scriptsize parameterized} & S & \scriptsize uses parameterized query \\
\code{\scriptsize sql query js} & \code{\scriptsize string concat sql} & I & \scriptsize concatenates user input into SQL \\
\code{\scriptsize sql query js} & \code{\scriptsize uses orm} & S & \scriptsize uses an ORM \\
\code{\scriptsize cors config js} & \code{\scriptsize wildcard origin} & I & \scriptsize allows any origin (*) \\
\code{\scriptsize cors config js} & \code{\scriptsize explicit origin} & S & \scriptsize restricts to explicit origins \\
\code{\scriptsize cors config js} & \code{\scriptsize credentials flag} & S & \scriptsize handles credentials explicitly \\
\code{\scriptsize file upload js} & \code{\scriptsize extension check} & S & \scriptsize validates file type/extension \\
\code{\scriptsize file upload js} & \code{\scriptsize secure filename} & S & \scriptsize sanitizes the filename \\
\code{\scriptsize file upload js} & \code{\scriptsize path traversal risk} & I & \scriptsize joins raw filename into a path \\
\code{\scriptsize file upload js} & \code{\scriptsize size limit} & S & \scriptsize limits upload size \\
\code{\scriptsize session cookie js} & \code{\scriptsize httponly} & S & \scriptsize sets HttpOnly \\
\code{\scriptsize session cookie js} & \code{\scriptsize secure flag} & S & \scriptsize sets Secure \\
\code{\scriptsize session cookie js} & \code{\scriptsize samesite} & S & \scriptsize sets SameSite \\
\code{\scriptsize session cookie js} & \code{\scriptsize hardcoded secret key} & I & \scriptsize hardcodes the session secret \\
\code{\scriptsize password hash go} & \code{\scriptsize strong kdf} & S & \scriptsize uses bcrypt/scrypt/argon2 \\
\code{\scriptsize password hash go} & \code{\scriptsize weak hash} & I & \scriptsize uses md5/sha1 \\
\code{\scriptsize password hash go} & \code{\scriptsize per user salt} & S & \scriptsize uses a salt \\
\code{\scriptsize password hash go} & \code{\scriptsize constant time cmp} & S & \scriptsize safe compare / library verify \\
\code{\scriptsize jwt verify go} & \code{\scriptsize verify signature} & S & \scriptsize verifies the signature \\
\code{\scriptsize jwt verify go} & \code{\scriptsize alg pinned} & S & \scriptsize pins allowed algorithms \\
\code{\scriptsize jwt verify go} & \code{\scriptsize alg none risk} & I & \scriptsize does not check signing method \\
\code{\scriptsize jwt verify go} & \code{\scriptsize exp checked} & S & \scriptsize checks expiry \\
\code{\scriptsize jwt verify go} & \code{\scriptsize uses library} & S & \scriptsize uses a JWT library \\
\code{\scriptsize sql query go} & \code{\scriptsize parameterized} & S & \scriptsize uses parameterized query \\
\code{\scriptsize sql query go} & \code{\scriptsize string concat sql} & I & \scriptsize builds SQL with Sprintf/concat \\
\code{\scriptsize sql query go} & \code{\scriptsize uses orm} & S & \scriptsize uses an ORM \\
\code{\scriptsize cors config go} & \code{\scriptsize wildcard origin} & I & \scriptsize allows any origin (*) \\
\code{\scriptsize cors config go} & \code{\scriptsize explicit origin} & S & \scriptsize restricts to explicit origins \\
\code{\scriptsize cors config go} & \code{\scriptsize credentials flag} & S & \scriptsize handles credentials explicitly \\
\code{\scriptsize file upload go} & \code{\scriptsize extension check} & S & \scriptsize validates file type/extension \\
\code{\scriptsize file upload go} & \code{\scriptsize secure filename} & S & \scriptsize sanitizes the filename \\
\code{\scriptsize file upload go} & \code{\scriptsize path traversal risk} & I & \scriptsize joins raw filename into a path \\
\code{\scriptsize file upload go} & \code{\scriptsize size limit} & S & \scriptsize limits upload size \\
\code{\scriptsize session cookie go} & \code{\scriptsize httponly} & S & \scriptsize sets HttpOnly \\
\code{\scriptsize session cookie go} & \code{\scriptsize secure flag} & S & \scriptsize sets Secure \\
\code{\scriptsize session cookie go} & \code{\scriptsize samesite} & S & \scriptsize sets SameSite \\
\code{\scriptsize session cookie go} & \code{\scriptsize hardcoded secret key} & I & \scriptsize hardcodes the session key \\
\code{\scriptsize oauth pkce go} & \code{\scriptsize pkce} & S & \scriptsize uses PKCE \\
\code{\scriptsize oauth pkce go} & \code{\scriptsize state param} & S & \scriptsize uses anti-CSRF state parameter \\
\code{\scriptsize oauth pkce go} & \code{\scriptsize redirect validation} & S & \scriptsize validates redirect uri \\
\code{\scriptsize oauth pkce go} & \code{\scriptsize https enforced} & S & \scriptsize uses https endpoints \\
\code{\scriptsize oauth pkce go} & \code{\scriptsize hardcoded secret} & I & \scriptsize hardcodes client secret \\
\code{\scriptsize oauth pkce go} & \code{\scriptsize uses library} & S & \scriptsize uses an OAuth library \\
\code{\scriptsize command exec} & \code{\scriptsize shell true risk} & I & \scriptsize uses shell=True with input \\
\code{\scriptsize command exec} & \code{\scriptsize arg list} & S & \scriptsize passes arguments as a list \\
\code{\scriptsize command exec} & \code{\scriptsize input validation} & S & \scriptsize validates/sanitizes input \\
\code{\scriptsize xss escape} & \code{\scriptsize auto escape} & S & \scriptsize uses a template/auto-escaping \\
\code{\scriptsize xss escape} & \code{\scriptsize raw html risk} & I & \scriptsize inserts input into HTML string directly \\
\code{\scriptsize xss escape} & \code{\scriptsize csp} & S & \scriptsize sets a content security policy \\
\code{\scriptsize ssrf fetch} & \code{\scriptsize scheme check} & S & \scriptsize validates the URL scheme \\
\code{\scriptsize ssrf fetch} & \code{\scriptsize host allowlist} & S & \scriptsize restricts the host \\
\code{\scriptsize ssrf fetch} & \code{\scriptsize no validation risk} & I & \scriptsize fetches the raw url with no checks \\
\code{\scriptsize secrets env} & \code{\scriptsize env or vault} & S & \scriptsize reads the secret from env/secret store \\
\code{\scriptsize secrets env} & \code{\scriptsize hardcoded key} & I & \scriptsize hardcodes the key literal \\
\code{\scriptsize path serve} & \code{\scriptsize safe join} & S & \scriptsize uses a safe path join / basename \\
\code{\scriptsize path serve} & \code{\scriptsize traversal risk} & I & \scriptsize joins the raw name into a path \\
\code{\scriptsize oauth2 spring} & \code{\scriptsize uses framework} & S & \scriptsize uses Spring Security OAuth2 client \\
\code{\scriptsize oauth2 spring} & \code{\scriptsize pkce} & S & \scriptsize enables PKCE \\
\code{\scriptsize oauth2 spring} & \code{\scriptsize csrf protection} & S & \scriptsize keeps CSRF protection \\
\code{\scriptsize oauth2 spring} & \code{\scriptsize csrf disabled} & I & \scriptsize disables CSRF protection \\
\code{\scriptsize oauth2 spring} & \code{\scriptsize hardcoded secret} & I & \scriptsize hardcodes the client secret \\
\code{\scriptsize oauth2 spring} & \code{\scriptsize externalized secret} & S & \scriptsize reads secret from config/env \\
\code{\scriptsize oauth2 django} & \code{\scriptsize uses framework} & S & \scriptsize uses a Django OAuth library \\
\code{\scriptsize oauth2 django} & \code{\scriptsize pkce} & S & \scriptsize enables PKCE \\
\code{\scriptsize oauth2 django} & \code{\scriptsize https callback} & S & \scriptsize uses https callback \\
\code{\scriptsize oauth2 django} & \code{\scriptsize hardcoded secret} & I & \scriptsize hardcodes the client secret \\
\code{\scriptsize oauth2 django} & \code{\scriptsize externalized secret} & S & \scriptsize reads secret from env/config \\
\code{\scriptsize oauth2 django} & \code{\scriptsize debug on} & I & \scriptsize ships with DEBUG True \\
\code{\scriptsize oauth2 express} & \code{\scriptsize uses framework} & S & \scriptsize uses passport oauth2 strategy \\
\code{\scriptsize oauth2 express} & \code{\scriptsize state param} & S & \scriptsize uses state parameter \\
\code{\scriptsize oauth2 express} & \code{\scriptsize session secret env} & S & \scriptsize reads session secret from env \\
\code{\scriptsize oauth2 express} & \code{\scriptsize hardcoded secret} & I & \scriptsize hardcodes the client secret \\
\code{\scriptsize oauth2 express} & \code{\scriptsize https callback} & S & \scriptsize uses https callback url \\
\code{\scriptsize spring jpa query} & \code{\scriptsize uses orm} & S & \scriptsize uses Spring Data repository \\
\code{\scriptsize spring jpa query} & \code{\scriptsize param binding} & S & \scriptsize uses parameter binding \\
\code{\scriptsize spring jpa query} & \code{\scriptsize string concat sql} & I & \scriptsize concatenates input into a query \\
\end{longtable}

\begin{longtable}{@{}p{3.4cm}p{3.4cm}rrrr@{}}
\caption{Complete per-subject, per-pattern adoption for the human (H) and model (L) corpora, with sample counts.}\label{tab:fullprev}\\
\toprule
Subject & Pattern & H adopt & H $n$ & L adopt & L $n$ \\
\midrule
\endfirsthead
\multicolumn{6}{c}{\footnotesize\itshape Table \thetable\ continued}\\ \toprule Subject & Pattern & H adopt & H $n$ & L adopt & L $n$ \\ \midrule
\endhead
\midrule \multicolumn{6}{r}{\footnotesize\itshape continued}\\
\endfoot
\bottomrule
\endlastfoot
\code{\scriptsize command exec} & \code{\scriptsize arg list} & 0.143 & 7 & 0.167 & 12 \\
\code{\scriptsize } & \code{\scriptsize input validation} & 0.0 & 7 & 0.25 & 12 \\
\code{\scriptsize } & \code{\scriptsize shell true risk} & 0.143 & 7 & 0.417 & 12 \\
\code{\scriptsize cors config} & \code{\scriptsize credentials flag} & 0.0 & 8 & 0.267 & 15 \\
\code{\scriptsize } & \code{\scriptsize explicit origin} & 0.375 & 8 & 0.6 & 15 \\
\code{\scriptsize } & \code{\scriptsize wildcard origin} & 0.5 & 8 & 0.133 & 15 \\
\code{\scriptsize cors config go} & \code{\scriptsize credentials flag} & 0.25 & 8 & 0.833 & 12 \\
\code{\scriptsize } & \code{\scriptsize explicit origin} & 0.25 & 8 & 0.833 & 12 \\
\code{\scriptsize } & \code{\scriptsize wildcard origin} & 0.125 & 8 & 0.417 & 12 \\
\code{\scriptsize cors config js} & \code{\scriptsize credentials flag} & 0.0 & 8 & 0.571 & 14 \\
\code{\scriptsize } & \code{\scriptsize explicit origin} & 0.25 & 8 & 0.714 & 14 \\
\code{\scriptsize } & \code{\scriptsize wildcard origin} & 0.125 & 8 & 0.429 & 14 \\
\code{\scriptsize file upload} & \code{\scriptsize extension check} & 0.125 & 8 & 0.267 & 15 \\
\code{\scriptsize } & \code{\scriptsize path traversal risk} & 0.0 & 8 & 0.0 & 15 \\
\code{\scriptsize } & \code{\scriptsize secure filename} & 0.0 & 8 & 0.4 & 15 \\
\code{\scriptsize } & \code{\scriptsize size limit} & 0.0 & 8 & 0.133 & 15 \\
\code{\scriptsize file upload go} & \code{\scriptsize extension check} & -- & -- & 0.5 & 12 \\
\code{\scriptsize } & \code{\scriptsize path traversal risk} & -- & -- & 0.667 & 12 \\
\code{\scriptsize } & \code{\scriptsize secure filename} & -- & -- & 0.667 & 12 \\
\code{\scriptsize } & \code{\scriptsize size limit} & -- & -- & 1.0 & 12 \\
\code{\scriptsize file upload js} & \code{\scriptsize extension check} & -- & -- & 0.5 & 14 \\
\code{\scriptsize } & \code{\scriptsize path traversal risk} & -- & -- & 0.071 & 14 \\
\code{\scriptsize } & \code{\scriptsize secure filename} & -- & -- & 0.143 & 14 \\
\code{\scriptsize } & \code{\scriptsize size limit} & -- & -- & 0.214 & 14 \\
\code{\scriptsize jwt verify} & \code{\scriptsize alg none risk} & 0.0 & 8 & 0.059 & 17 \\
\code{\scriptsize } & \code{\scriptsize alg pinned} & 0.375 & 8 & 0.882 & 17 \\
\code{\scriptsize } & \code{\scriptsize exp checked} & 0.25 & 8 & 0.882 & 17 \\
\code{\scriptsize } & \code{\scriptsize uses library} & 0.5 & 8 & 1.0 & 17 \\
\code{\scriptsize } & \code{\scriptsize verify signature} & 0.625 & 8 & 0.941 & 17 \\
\code{\scriptsize jwt verify go} & \code{\scriptsize alg none risk} & 0.0 & 8 & 0.0 & 12 \\
\code{\scriptsize } & \code{\scriptsize alg pinned} & 0.5 & 8 & 0.833 & 12 \\
\code{\scriptsize } & \code{\scriptsize exp checked} & 0.375 & 8 & 0.833 & 12 \\
\code{\scriptsize } & \code{\scriptsize uses library} & 0.375 & 8 & 0.917 & 12 \\
\code{\scriptsize } & \code{\scriptsize verify signature} & 0.75 & 8 & 0.917 & 12 \\
\code{\scriptsize jwt verify js} & \code{\scriptsize alg none risk} & 0.0 & 4 & 0.0 & 14 \\
\code{\scriptsize } & \code{\scriptsize alg pinned} & 0.0 & 4 & 0.071 & 14 \\
\code{\scriptsize } & \code{\scriptsize exp checked} & 0.25 & 4 & 0.929 & 14 \\
\code{\scriptsize } & \code{\scriptsize uses library} & 0.25 & 4 & 0.929 & 14 \\
\code{\scriptsize } & \code{\scriptsize verify signature} & 0.5 & 4 & 1.0 & 14 \\
\code{\scriptsize landing page} & \code{\scriptsize alt text} & -- & -- & 0.0 & 13 \\
\code{\scriptsize } & \code{\scriptsize flex or grid} & -- & -- & 0.769 & 13 \\
\code{\scriptsize } & \code{\scriptsize inline style} & -- & -- & 0.077 & 13 \\
\code{\scriptsize } & \code{\scriptsize media query} & -- & -- & 0.923 & 13 \\
\code{\scriptsize } & \code{\scriptsize semantic html} & -- & -- & 1.0 & 13 \\
\code{\scriptsize } & \code{\scriptsize viewport meta} & -- & -- & 1.0 & 13 \\
\code{\scriptsize oauth2 django} & \code{\scriptsize debug on} & 0.0 & 2 & 0.0 & 12 \\
\code{\scriptsize } & \code{\scriptsize externalized secret} & 0.0 & 2 & 0.25 & 12 \\
\code{\scriptsize } & \code{\scriptsize hardcoded secret} & 0.0 & 2 & 0.25 & 12 \\
\code{\scriptsize } & \code{\scriptsize https callback} & 0.5 & 2 & 0.0 & 12 \\
\code{\scriptsize } & \code{\scriptsize pkce} & 0.0 & 2 & 0.25 & 12 \\
\code{\scriptsize } & \code{\scriptsize uses framework} & 1.0 & 2 & 1.0 & 12 \\
\code{\scriptsize oauth2 express} & \code{\scriptsize hardcoded secret} & 0.0 & 3 & 0.167 & 12 \\
\code{\scriptsize } & \code{\scriptsize https callback} & 0.0 & 3 & 0.083 & 12 \\
\code{\scriptsize } & \code{\scriptsize session secret env} & 0.0 & 3 & 0.75 & 12 \\
\code{\scriptsize } & \code{\scriptsize state param} & 0.0 & 3 & 0.25 & 12 \\
\code{\scriptsize } & \code{\scriptsize uses framework} & 0.0 & 3 & 1.0 & 12 \\
\code{\scriptsize oauth2 spring} & \code{\scriptsize csrf disabled} & 0.125 & 8 & 0.0 & 12 \\
\code{\scriptsize } & \code{\scriptsize csrf protection} & 0.125 & 8 & 0.25 & 12 \\
\code{\scriptsize } & \code{\scriptsize externalized secret} & 0.125 & 8 & 0.5 & 12 \\
\code{\scriptsize } & \code{\scriptsize hardcoded secret} & 0.0 & 8 & 0.417 & 12 \\
\code{\scriptsize } & \code{\scriptsize pkce} & 0.0 & 8 & 0.0 & 12 \\
\code{\scriptsize } & \code{\scriptsize uses framework} & 0.875 & 8 & 1.0 & 12 \\
\code{\scriptsize oauth pkce} & \code{\scriptsize hardcoded secret} & 0.0 & 7 & 0.333 & 18 \\
\code{\scriptsize } & \code{\scriptsize https enforced} & 0.571 & 7 & 0.778 & 18 \\
\code{\scriptsize } & \code{\scriptsize pkce} & 0.0 & 7 & 1.0 & 18 \\
\code{\scriptsize } & \code{\scriptsize redirect validation} & 0.0 & 7 & 0.0 & 18 \\
\code{\scriptsize } & \code{\scriptsize state param} & 0.143 & 7 & 0.667 & 18 \\
\code{\scriptsize } & \code{\scriptsize uses library} & 0.571 & 7 & 0.056 & 18 \\
\code{\scriptsize oauth pkce go} & \code{\scriptsize hardcoded secret} & -- & -- & 0.333 & 12 \\
\code{\scriptsize } & \code{\scriptsize https enforced} & -- & -- & 0.667 & 12 \\
\code{\scriptsize } & \code{\scriptsize pkce} & -- & -- & 1.0 & 12 \\
\code{\scriptsize } & \code{\scriptsize redirect validation} & -- & -- & 0.0 & 12 \\
\code{\scriptsize } & \code{\scriptsize state param} & -- & -- & 0.75 & 12 \\
\code{\scriptsize } & \code{\scriptsize uses library} & -- & -- & 0.417 & 12 \\
\code{\scriptsize oauth pkce js} & \code{\scriptsize hardcoded secret} & -- & -- & 0.643 & 14 \\
\code{\scriptsize } & \code{\scriptsize https enforced} & -- & -- & 0.786 & 14 \\
\code{\scriptsize } & \code{\scriptsize pkce} & -- & -- & 1.0 & 14 \\
\code{\scriptsize } & \code{\scriptsize redirect validation} & -- & -- & 0.0 & 14 \\
\code{\scriptsize } & \code{\scriptsize state param} & -- & -- & 0.357 & 14 \\
\code{\scriptsize } & \code{\scriptsize uses library} & -- & -- & 0.0 & 14 \\
\code{\scriptsize password hash} & \code{\scriptsize constant time cmp} & 0.0 & 6 & 0.733 & 15 \\
\code{\scriptsize } & \code{\scriptsize per user salt} & 0.333 & 6 & 0.933 & 15 \\
\code{\scriptsize } & \code{\scriptsize strong kdf} & 0.167 & 6 & 1.0 & 15 \\
\code{\scriptsize } & \code{\scriptsize weak hash} & 0.167 & 6 & 0.0 & 15 \\
\code{\scriptsize password hash go} & \code{\scriptsize constant time cmp} & -- & -- & 0.917 & 12 \\
\code{\scriptsize } & \code{\scriptsize per user salt} & -- & -- & 0.917 & 12 \\
\code{\scriptsize } & \code{\scriptsize strong kdf} & -- & -- & 1.0 & 12 \\
\code{\scriptsize } & \code{\scriptsize weak hash} & -- & -- & 0.0 & 12 \\
\code{\scriptsize password hash js} & \code{\scriptsize constant time cmp} & -- & -- & 0.786 & 14 \\
\code{\scriptsize } & \code{\scriptsize per user salt} & -- & -- & 0.929 & 14 \\
\code{\scriptsize } & \code{\scriptsize strong kdf} & -- & -- & 1.0 & 14 \\
\code{\scriptsize } & \code{\scriptsize weak hash} & -- & -- & 0.0 & 14 \\
\code{\scriptsize path serve} & \code{\scriptsize safe join} & 0.0 & 1 & 0.667 & 12 \\
\code{\scriptsize } & \code{\scriptsize traversal risk} & 0.0 & 1 & 0.417 & 12 \\
\code{\scriptsize secrets env} & \code{\scriptsize env or vault} & 0.0 & 5 & 1.0 & 12 \\
\code{\scriptsize } & \code{\scriptsize hardcoded key} & 0.0 & 5 & 0.0 & 12 \\
\code{\scriptsize session cookie} & \code{\scriptsize hardcoded secret key} & 0.0 & 3 & 0.786 & 14 \\
\code{\scriptsize } & \code{\scriptsize httponly} & 0.0 & 3 & 0.571 & 14 \\
\code{\scriptsize } & \code{\scriptsize samesite} & 0.0 & 3 & 0.5 & 14 \\
\code{\scriptsize } & \code{\scriptsize secure flag} & 0.0 & 3 & 0.857 & 14 \\
\code{\scriptsize session cookie go} & \code{\scriptsize hardcoded secret key} & -- & -- & 0.083 & 12 \\
\code{\scriptsize } & \code{\scriptsize httponly} & -- & -- & 1.0 & 12 \\
\code{\scriptsize } & \code{\scriptsize samesite} & -- & -- & 0.667 & 12 \\
\code{\scriptsize } & \code{\scriptsize secure flag} & -- & -- & 0.833 & 12 \\
\code{\scriptsize session cookie js} & \code{\scriptsize hardcoded secret key} & 0.0 & 1 & 0.571 & 14 \\
\code{\scriptsize } & \code{\scriptsize httponly} & 0.0 & 1 & 1.0 & 14 \\
\code{\scriptsize } & \code{\scriptsize samesite} & 0.0 & 1 & 0.643 & 14 \\
\code{\scriptsize } & \code{\scriptsize secure flag} & 0.0 & 1 & 0.143 & 14 \\
\code{\scriptsize spring jpa query} & \code{\scriptsize param binding} & 0.375 & 8 & 0.333 & 12 \\
\code{\scriptsize } & \code{\scriptsize string concat sql} & 0.0 & 8 & 0.0 & 12 \\
\code{\scriptsize } & \code{\scriptsize uses orm} & 0.0 & 8 & 1.0 & 12 \\
\code{\scriptsize sql query} & \code{\scriptsize parameterized} & 0.5 & 8 & 0.929 & 14 \\
\code{\scriptsize } & \code{\scriptsize string concat sql} & 0.25 & 8 & 0.0 & 14 \\
\code{\scriptsize } & \code{\scriptsize uses orm} & 0.375 & 8 & 0.0 & 14 \\
\code{\scriptsize sql query go} & \code{\scriptsize parameterized} & -- & -- & 0.75 & 12 \\
\code{\scriptsize } & \code{\scriptsize string concat sql} & -- & -- & 0.0 & 12 \\
\code{\scriptsize } & \code{\scriptsize uses orm} & -- & -- & 0.25 & 12 \\
\code{\scriptsize sql query js} & \code{\scriptsize parameterized} & -- & -- & 0.929 & 14 \\
\code{\scriptsize } & \code{\scriptsize string concat sql} & -- & -- & 0.0 & 14 \\
\code{\scriptsize } & \code{\scriptsize uses orm} & -- & -- & 0.0 & 14 \\
\code{\scriptsize ssrf fetch} & \code{\scriptsize host allowlist} & 0.5 & 2 & 0.333 & 12 \\
\code{\scriptsize } & \code{\scriptsize no validation risk} & 0.0 & 2 & 0.5 & 12 \\
\code{\scriptsize } & \code{\scriptsize scheme check} & 0.0 & 2 & 0.25 & 12 \\
\code{\scriptsize xss escape} & \code{\scriptsize auto escape} & 0.0 & 4 & 1.0 & 12 \\
\code{\scriptsize } & \code{\scriptsize csp} & 0.0 & 4 & 0.0 & 12 \\
\code{\scriptsize } & \code{\scriptsize raw html risk} & 0.25 & 4 & 0.0 & 12 \\
\end{longtable}

\twocolumn

\balance
\bibliographystyle{IEEEtran}
\bibliography{tables/refs}

\end{document}